\documentclass[
twocolumn,f
aps,prd,
nofootinbib,
superscriptaddress,
showpacs,
tightenlines,
]{revtex4}
\usepackage{amsmath}
\usepackage{amssymb}
\usepackage{bm}
\usepackage{color,graphicx}

\begin{document}

\title{Flexible Parametrization of Generalized Parton Distributions from Deeply Virtual Compton Scattering 
Observables}

\author{Gary R.~Goldstein} 
\email{gary.goldstein@tufts.edu}
\affiliation{Department of Physics and Astronomy, Tufts University, Medford, MA 02155 USA.}

\author{J. Osvaldo Gonzalez Hernandez} 
\email{jog4m@virginia.edu}
\affiliation{Department of Physics, University of Virginia, Charlottesville, VA 22901, USA.}

\author{Simonetta Liuti} 
\email{sl4y@virginia.edu}
\affiliation{Department of Physics, University of Virginia, Charlottesville, VA 22901, USA.}

\pacs{13.60.Hb, 13.40.Gp, 24.85.+p}

\begin{abstract}
We present a physically motivated parametrization of the chiral-even
generalized parton distributions in the non-singlet sector obtained from 
a global analysis using a set of available experimental data.
Our analysis is valid in the kinematical region of intermediate Bjorken $x$ and for 
$Q^2$  in the multi-GeV region which is accessible at present and currently planned facilities.
Relevant data included in our fit are from the nucleon elastic form factors measurements, 
from deep inelastic scattering experiments.
Additional information provided by lattice 
calculations of the higher moments of generalized parton distributions, is also
considered.
Recently extracted observables from Deeply Virtual Compton Scattering
on the nucleon are reproduced by our fit.
\end{abstract}

\maketitle

\section{Introduction}
High energy exclusive leptoproduction processes  have been drawing increasing attention after a long hiatus since they were first proposed as direct probes of  partonic structure.
%
The first  exclusive electron proton scattering experiments were conducted at both DESY
(H1, ZEUS and HERMES) and Jefferson Lab. A new forthcoming dedicated set of experiments are currently being performed and 
planned at both Jefferson Lab \cite{Jlab_exp} and CERN (Compass) \cite{Compass_exp}. The possibility of using neutrino beams to study hard exclusive reactions
also concretely exists  within {\it e.g.} the Minerva experiment at Fermilab \cite{Minerva_exp}. 

\begin{figure}
\includegraphics[width=9.cm]{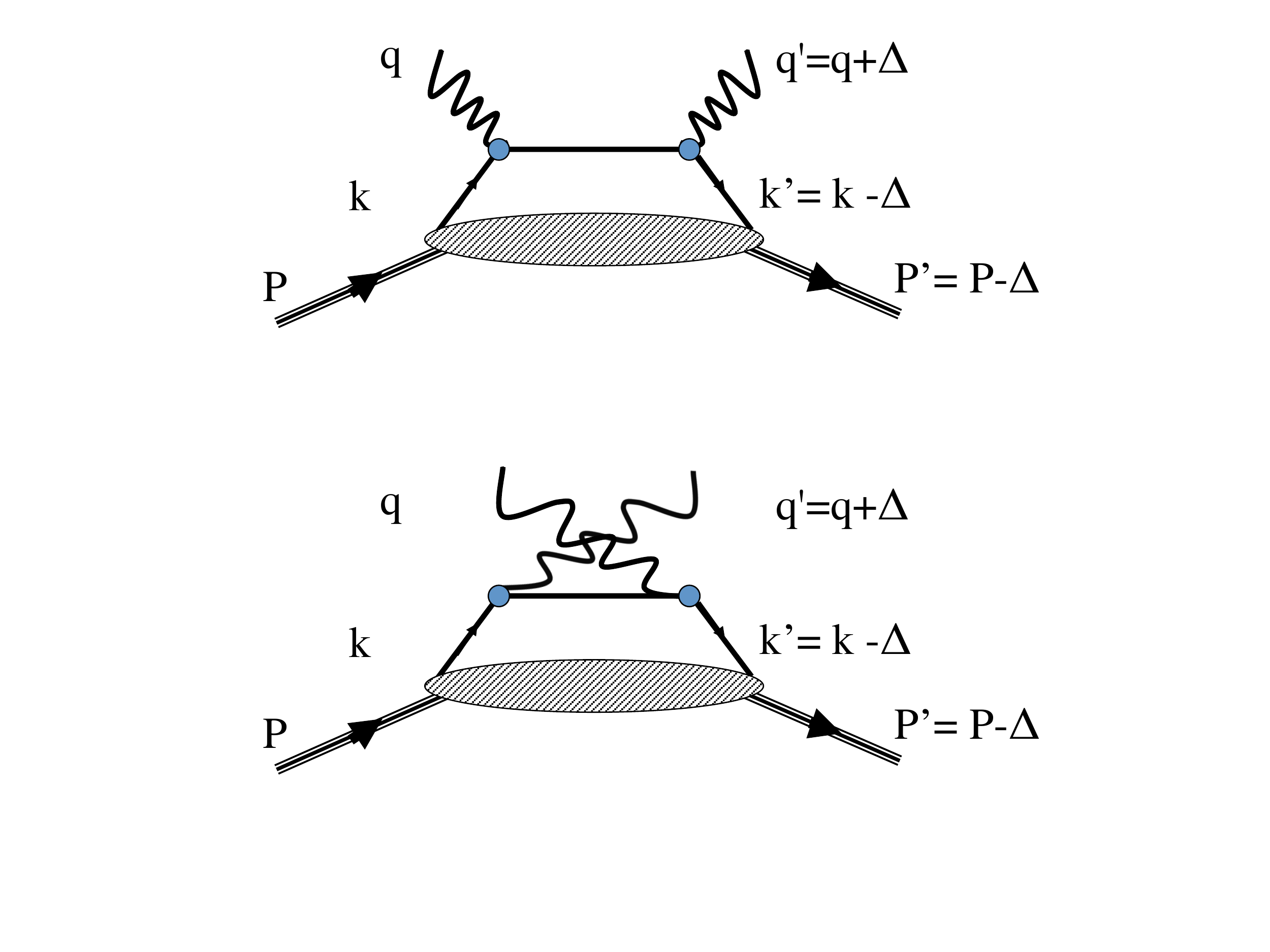}
\caption{Leading order amplitude for the DVCS process, $\gamma^* + P \rightarrow \gamma +P^\prime$}
\label{fig1}
\end{figure}

The interest  in deeply virtual exclusive processes originates from the realization first discussed in Refs.\cite{DMul1,Ji1,Rad1}
that  QCD factorization theorems similar to the inclusive DIS case can be proven.  
Collinear factorization theorems have in fact so far been established for Deeply Virtual Compton Scattering (DVCS), involving initially transverse photons, and for Deeply Virtual  
Meson Production (DVMP), with initial longitudinal photons \cite{CFS}. New developments in QCD factorization are also rapidly evolving  \cite{Collins}.  
The leading order diagrams describing the amplitude for the scattering process are shown in 
Figure \ref{fig1}. A phenomenology ensues
similar to the one extensively developed for inclusive scattering, an important difference being that exclusive reactions provide us with additional kinematical dependence on the momentum transfer squared 
between the initial and final proton, $t$, and on its Light Cone (LC) component, $\zeta$.
The new t-channel variables allow us to pin down in principle the dependence of the parton distributions on spatial d.o.f. through  Fourier transforms of the Generalized Parton Distributions (GPDs) \cite{Bur}.  The latter enter the description of the soft matrix elements for DVCS, DVMP and related processes.
%
The kinematical variables external to the partonic loop in Fig.\ref{fig1}a are the initial photon's
virtuality, $Q^2$, the skewness, $\zeta= (\Delta q)/(Pq) \simeq Q^2/2(Pq) \equiv x_{Bj}$, $t=\Delta^2$, with $\Delta=P-P^\prime$, $P$ ($P^\prime$) being the
initial (final) photon momentum.  $(\zeta,t,Q^2)$ define a set of independent invariants. 
In the factorized approach one defines also internal loop  variables, $X=(kq)/(Pq)$ -- the parton's momentum fraction -- and $k_T$ -- the intrinsic transverse momentum. 

At high momentum transfer, the amplitude for DVCS can be written schematically as
\begin{equation}
\label{DVCS}
T^{\mu\nu}(\zeta,t,Q^2)=\frac{1}{2}g^{\mu\nu}_T{\overline U}(P^\prime){\not\!n}U(P)
\sum_{q} e_q^2 {\cal H}_q (\zeta,t,Q^2),
\end{equation}
where we considered for ease of presentation only the GPD $H$ (detailed expressions will be given in what follows). The analog of the Compton Form Factor (CFF) with one virtual photon is
\begin{eqnarray}
\label{direct}
{\cal H}_q(\zeta,t,Q^2) & = & \int\limits_{-1+\zeta}^{+1}dX  H_q (X,\zeta,t,Q^2) \nonumber \\%
& & \times \left( \frac{1}{X-\zeta+i\epsilon} + \frac{1}{X- i \epsilon}  \right).
\end{eqnarray}
One has therefore that both the imaginary and real parts of the amplitude, namely
\begin{subequations}
\begin{eqnarray}
{\rm Im} \,  \mathcal{H}_q(\zeta,t) & = & \pi [ H_q(\zeta,\zeta,t) - H_q(0,\zeta,t) ]
\label{Im_H}
\\
{\rm Re} \, {\cal H}_q(\zeta,t) &  =  &  P.V. \int_{-1}^{1} dX   H_q(X,\zeta,t)  \nonumber \\
& & \times\left( \frac{1}{X-\zeta+i\epsilon} + \frac{1}{X - i \epsilon}  \right),  
\label{Re_H}
\end{eqnarray}
\end{subequations}
enter the description the DVCS reaction. 
Information on the partonic distributions which is contained  $H_q(X,\zeta,t)$, 
needs to be extracted from these observables.
This is an important difference with DIS where, because of the optical theorem, the cross section by definition measures the imaginary
part of the forward amplitude. The DIS cross section is therefore directly proportional to linear combinations of the soft matrix elements, or Parton Distributions Functions (PDFs) convoluted with appropriate Wilson coefficient functions. 
On the contrary, in DVCS, DVMP, and related processes one needs to disentangle both the real and imaginary contributions of the CFFs defined in Eqs.(\ref{Im_H},\ref{Re_H}) \cite{BKM}.

It was recently suggested to use dispersion relations in order to relate the real and imaginary parts of CFFs. However, as we noted in \cite{GolLiu}, dispersion relations do not apply straightforwardly because of the appearance of $t$-dependent physical thresholds. 
As a result, we reiterate that both the real and imaginary parts need to be extracted separately from experiment,  at variance with the simplification suggested  {\it e.g.} in Refs.\cite{AniTer,DieIva}  . 

On one hand, the type of information we wish to obtain from high energy exclusive experiments  is
a sufficiently large range of GPD values in $(\zeta,t,Q^2)$ that would enable us to reconstruct the partonic spatial distributions of the nucleon from a Fourier transformation in ${\bf \Delta}_\perp$.
This  would allow us both to explore the holographic principle for the nucleon, and to connect to Transverse Momentum Distributions (TMDs).  On the other hand,  it is important to have
access  to the spin flip GPD, $E$, which is essential  for determining the Orbital Angular Momentum (OAM) contribution to the spin sum rule .

The question of whether the various GPDs can be extracted reliably from current experiments has been raised, given the complications
inherent both in their convolution form, and in their complex multi-variable analysis  (see {\it e.g.} \cite{KumMuel1,KumMuel2}). A pragmatic response was given in \cite{GuiMou,Mou} 
where 
an assessment was made of which GPDs can be extracted using the present body of data  from Jefferson Lab and Hermes. In particular, it was concluded that the only CFFs that are presently constrained by experiments are ${\rm Re}{\cal H}$ and ${\rm Im}{\cal H}$, with rather large errors, up to $30\%$.   A global fit using the dual model of Ref.\cite{KumMuel1}, valid mostly at low Bjorken $x$ was also conducted in \cite{KumMuel2}.
However, these approaches raise many concerns. In particular,  can the "dual model" used in the fits accommodate all of the data with the given number of parameters?
The problem is 
critical, in particular,  for both higher $\zeta$ values and for the real CFFs.
Furthermore, the analysis of \cite{Mou} does obtain model independent extractions of CFFs at the 
expense of not allowing for extrapolations to kinematical domains beyond the very sparse data sets.  


The goal of our fit is to extract the GPDs from a variety of experiments
under the following basic assumptions:

\noindent {\it i)} QCD Factorization is working, namely the soft and hard parts are separated as shown in Fig.\ref{fig1};

\noindent {\it ii)} the GPDs contributing to DVCS are evaluated at the lowest order in QCD.
  
This situation is somewhat similar to the extraction of PDFs from structure functions at NLO, where the PDFs are convoluted 
with the NLO Wilson coefficient functions. 
The convolution is neither a substantial or conceptual obstacle so long as one is providing an appropriate initial functional shape. 
The strategy we propose here provides a parametric form of the chiral even GPDs, $H, E, \widetilde{H}, \widetilde{E}$ that is valid in the multi-GeV, intermediate $x_{Bj}$ region accessible at Jefferson Lab and COMPASS. 

We suggest the idea that for extracting GPDs from experiment a {\em progressive/recursive fit} 
should be used rather than a global fit. In our fitting procedure constraints are applied sequentially, the final result being updated upon including each new constraint.
In a nutshell, in a first step we provide a flexible form that includes all constraints from inclusive data --  DIS structure functions and elastic electroweak form factors. We subsequently evaluate the impact of presently available DVCS data from both Jefferson Lab and HERMES. The data set used in our analysis is consistent with the one from Ref.\cite{GuiMou}. 
The parametric form is based on a ``Regge improved" diquark model that because of its similarities and possible connections with the dual model \cite{Pol}, we call the {\it hybrid model}.

Our approach however makes two important distinctions: 
{\it i)}  we attack the GPD parametrization issue from the bottom-up perspective. We adopt a flexible parametrization that is consistent with theoretical constraints imposed numerically, and let the experimental data guide the shape of the parametrization as closely as possible.
In this procedure, experimental evidence is used to constrain the various theoretical aspects of the GPDs behavior, eventually giving rise to a complete model; 
 {\it ii)} our model differs from some of the lore on the partonic interpretation of GPDs in the ERBL region. In Ref.\cite{GL_partons} we in fact pointed out that the ERBL region, or the region with $X<\zeta$, cannot be described in terms of a quark anti-quark pair emerging from the nucleon because of the presence of semi-disconnected, unphysical, diagrams associated to this configuration (Fig.\ref{fig1}b). While casting a doubt  on any partonic picture in the ERBL region, we also suggested that  multi-parton 
distributions may restore the connectedness, and consequently the partonic  interpretation of the DVCS graphs. 
In this paper we therefore adopted, as a practical scheme, the hybrid model in the DGLAP ($X>\zeta$) region, and a minimal model that is consistent with the properties of continuity at $X=\zeta$, polynomiality, and crossing symmetry for the ERBL region.  

Details on the model are given in Section \ref{sec2}. 
The new fitting procedure is described in Section \ref{sec3}. 
In Section \ref{sec4}  
we discuss the implementation of world DVCS data. 
Finally, in Section \ref{sec5} we draw our conclusions and outline future work. 

\section{Covariant formulation and symmetries}
\label{sec2}
We begin by describing the connection between the Dirac basis formulation of the correlation function and the helicity amplitudes formalism in DVCS.
Some of this formalism was outlined in Refs.\cite{Diehl_rev,AGL}. We, however, present here the formal details that will be important for the extraction of observables in 
Section \ref{sec4}.   

\subsection{Formalism}
The factorization theorem for hard scattering processes 
allows us to separate the hard scattering between the elementary constituents, which is calculated using perturbative QCD, 
from the soft hadronic matrix element, ${\cal M}$, as
\begin{eqnarray}
T^{\mu \nu} &  = &  -i \int \frac{d^4 k}{(2\pi)^4} 
{\rm Tr} \left[ \left(
\frac{\gamma^\mu i (\not\!k + \not\!q)\gamma^\nu}{(k+q)^2+i\epsilon}  \right. \right. \nonumber \\
& + &
\left. \left. \frac{\gamma^\nu i(\not\!k - \not\!\Delta - \not\!q)\gamma^\mu}{(k-\Delta-q)^2+i\epsilon} \right) \, {\cal M}(k,P,\Delta) \right]. 
\label{h_tensor}
\end{eqnarray}
${\cal M}(k,P,\Delta)$ is the off-forward correlation function:
\begin{eqnarray}
{\cal M}_{ij}^{\Lambda \Lambda^\prime} (k,P,\Delta)=\int  d^4 y  \, {e^{iky}}
\langle P^\prime, \Lambda^\prime | \overline{\psi}_{j}(0)\psi_{i}(y) | P,  \Lambda \rangle,
\label{matrix}
\end{eqnarray}
where we have written out explicitly both the Dirac indices $i,j$, and the target's spins $\Lambda, \Lambda^\prime$. 
By projecting out the
dominant contribution 
in the Bjorken limit ($Q^2 \rightarrow \infty$, $x_{Bj} = Q^2/2M \nu \approx \zeta$ and $t$ fixed), which corresponds to transverse virtual photon polarization, one obtains
\begin{equation}
T^{\mu\nu} = \frac{1}{2} g^{\mu \nu}_T  {\cal F}_S^{\Lambda \Lambda^\prime} + \frac{i}{2} \epsilon^{\mu \nu}_T  {\cal F}_A^{\Lambda \Lambda^\prime}
\label{tensor_trans}
\end{equation}
where
$\displaystyle  g^{\mu\nu} _T = g^{\mu\nu} - p^\mu n^\nu -  p^\nu n^\mu$,  
$\displaystyle  \epsilon^{\mu\nu} _T = \epsilon_{\alpha \beta \sigma \rho} g^{\alpha\mu} _T g^{\beta\nu} _T n^\rho 
p^\sigma \equiv \epsilon^{-+\mu\nu}$, $p$ and $n$ being unit light cone vectors.
The labels $S$ and $A$ refer to the symmetric and antisymmetric components of the hadronic tensor with respect to $\mu \leftrightarrow \nu$, that will enter the unpolarized and longitudinally  polarized scattering, respectively, as will be clarified in what follows. 

A possible kinematical choice is the one where the struck quark's light cone (LC) longitudinal component is $k^+ = XP^+$, 
the momentum transfer, $\Delta_\mu$  is decomposed into a LC longitudinal component, $\Delta^+ = \zeta P^+$ and a 
transverse component, $\Delta_\perp$, such that the invariant $t=\Delta^2$ reads: 
$\displaystyle t= -\zeta^2 M^2 /(1-\zeta) - \frac{\Delta_\perp^2}{1-\zeta}$.  
\footnote{The formal difference between the so-called symmetric and asymmetric notations is explained {\it e.g.} in 
Ref.(\protect\cite{Diehl_rev}). }
\begin{figure}
\includegraphics[width=10.cm]{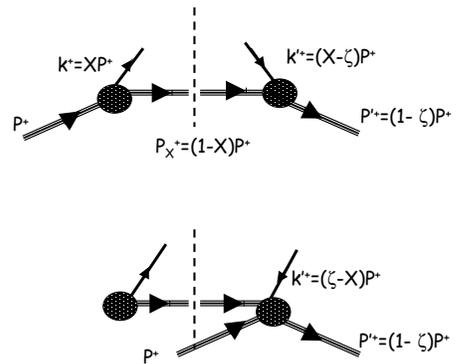}
\caption{Left: Amplitude for DVCS at leading order in $Q^2$. The light cone
components of the momenta for the active 
quarks and nucleons are explicitly written; 
Right: Time ordered diagrams for DVCS: {\bf (a)} dominant contribution in $X>\zeta$ region;
{\bf (b)} a $q \overline{q}$ pair is first produced from the nucleon and subsequently 
interacts with the photons. This process dominates the $X<\zeta$ region.
The crossed-terms where 
two of the particles in the same class are switched, are not shown in the figure.}
\label{fig2}
\end{figure}
Using these variables one can perform an integration over the quark loop momentum namely $d^4 k \equiv d k^+ d k^- d^2 k_\perp \equiv P^+ dX dk^- d^2 k_\perp$, obtaining the following expressions for the CFFs,
\begin{eqnarray}
{\cal F}^S_{\Lambda, \Lambda^\prime}(\zeta,t)  &=  &
\int_{-1+\zeta}^1 dX \left(\frac{1}{X- \zeta + i \epsilon } + 
\frac{1}{X - i \epsilon } \right)  \nonumber \\
& \times & F^S_{\Lambda, \Lambda^\prime}(X,\zeta,t) ,
\label{CFF}
\end{eqnarray}
\begin{eqnarray}
{\cal F}^A_{\Lambda, \Lambda^\prime}(\zeta,t) & =   &
\int_{-1+\zeta}^1 dX \left(-\frac{1}{X- \zeta + i \epsilon } + 
\frac{1}{X - i \epsilon } \right) \nonumber \\
& \times & F^A_{\Lambda, \Lambda^\prime}(X,\zeta,t), 
\label{CFFA}
\end{eqnarray}
where
\begin{widetext}
\begin{equation} 
F^S_{\Lambda, \Lambda^\prime}(X,\zeta,t) =
\frac{1}{2 \overline{P}^+} \left[ {\overline{U}(P',\Lambda')}\left( \gamma^+ H(X,\zeta,t)+
\frac{i \sigma^{+ \mu}(- \Delta_\mu)}{2M} E(X,\zeta,t) \right)U(P,\Lambda) \right],
\label{sfn}
\end{equation}
\begin{equation} 
F^A_{\Lambda, \Lambda^\prime}(X,\zeta,t) =
\frac{1}{2 \overline{P}^+} \left[ {\overline{U}(P',\Lambda')}\left( \gamma^+ \gamma_5 \widetilde{H}(X,\zeta,t)+
\gamma_5 \frac{- \Delta^+}{2M} \widetilde{E}(X,\zeta,t) \right)U(P,\Lambda) \right].
\label{sfnA}
\end{equation}
\end{widetext}
The  GPDs, $H, E, \widetilde{H}, \widetilde{E}$ introduced  in the
equations include the integration over $dk^- d^2 k_\perp$; 
$\overline{P}^+ = (P^+ + P^{+ \, \prime})/2$, and we 
did not write explicitly, for simplicity, both the label for the different quark components, and the dependence on the scale of the process, $Q^2$. 

Eqs.(\ref{sfn}) and (\ref{sfnA}) define the basic Dirac structure for the chiral-even sector at leading order in $1/Q$.  
The connection with the helicity formalism is made by introducing the helicity amplitudes for DVCS,
\begin{eqnarray}
f_{\Lambda_\gamma, \Lambda;  \Lambda_\gamma^\prime, \Lambda^\prime} = \epsilon_\mu^{\Lambda_\gamma} T^{\mu\nu}_{\Lambda \Lambda^\prime} \epsilon_\nu^{* \Lambda^{\prime}_\gamma}, 
\end{eqnarray}
where $\epsilon_\mu^\Lambda$, are the photon polarization vectors, $(\Lambda_\gamma, \Lambda)$ 
refer to the initial (virtual) photon and proton helicities, and $(\Lambda_\gamma^\prime, \Lambda^\prime)$ to the final ones.
The following 
decomposition of $f_{\Lambda_\gamma, \Lambda;  \Lambda_\gamma^\prime, \Lambda^\prime}$ \cite{AGL} can be made
\begin{eqnarray}
f_{\Lambda_\gamma, \Lambda;  \Lambda_\gamma^\prime, \Lambda^\prime} & =   & \sum_{\lambda,\lambda^\prime}  \, 
 g_{\lambda,\lambda^\prime}^{\Lambda_\gamma, \Lambda_\gamma^\prime} (X,\zeta,t; Q^2)  
 \otimes A_{\Lambda^\prime,\lambda^\prime;\Lambda,\lambda}(X,\zeta,t), \nonumber \\
\label{facto}
\end{eqnarray}
where
$ g_{\lambda,\lambda^\prime}^{\Lambda_\gamma, \Lambda_\gamma^\prime}$ describes the partonic subprocess $\gamma^* + q \rightarrow \gamma + q$, {\it i.e.}  the scattering of  a transverse virtual photon 
from a quark with polarization $\lambda$; $A_{\Lambda^\prime,\lambda^\prime;\Lambda,\lambda}$ is the quark-proton helicity amplitude, and the convolution integral
 is given by $\otimes \rightarrow \int_{-\zeta+1}^1 d X$.
 %
In the Bjorken limit $g_{\lambda,\lambda^\prime}^{\Lambda_\gamma, \Lambda_\gamma^\prime}$ reads
\begin{eqnarray} 
g_{\lambda,\lambda^\prime}^{\Lambda_\gamma, \Lambda_\gamma^\prime}(X,\zeta)&  =  & \left[ \bar{u}(k^\prime,\lambda^\prime) \gamma^\mu \gamma^+ \gamma^\nu u(k,\lambda) \right]  \nonumber \\ 
& \times & \left(\frac{ \epsilon_\mu^{\Lambda_\gamma} \epsilon_\nu^{* \: \Lambda_\gamma^\prime}}{\hat{s} - i \epsilon } + \frac{ \epsilon_\mu^{*\;\Lambda_\gamma^\prime} \epsilon_\nu^{ \Lambda_\gamma}}{\hat{u} -i \epsilon} \right) \, q^- 
\label{ampg}
\end{eqnarray}
with $\hat{s} = (k+q)^2 \approx Q^2(X-\zeta)/\zeta $ and $\hat{u} = (k^\prime -q)^2 \approx Q^2 X/\zeta$, and $q^- \approx(Pq)/P^+ = Q^2/(2\zeta P^+)$.
%
%

For DVCS one can consider either the sum over the transverse helicities of the initial photon, {\it i.e.} we take it as unpolarized, or the difference of the helicities in which case the transverse photon is polarized.  
The outgoing photon is on-shell, thus purely transverse or helicity $\pm 1$. So the leading incoming virtual photon will have the same helicity in the collinear limit. Only $g_{+,+}^{+,+} (=g_{-,-}^{-,-}$ via Parity conservation) for the direct $\hat{s}$ pole term or $g_{+,+}^{-,-} (=g_{-,-}^{+,+})$ for the crossed $\hat{u}$ pole term will survive. 
For either allowed combination,  $g_{++}^S = g_{++}^{++} + g_{++}^{--} $, and $g_{++}^A = g_{++}^{++} - g_{++}^{--} $,
\begin{eqnarray}
\label{ampg++}
g_{++}^{++} \pm g_{++}^{--}  & = & \sqrt{X(X-\zeta)}  \left(\frac{1}{X- \zeta + i \epsilon } \pm 
\frac{1}{X - i \epsilon } \right). \nonumber \\
\end{eqnarray}
where we use the $S$/$A$ labels  for the sum/difference between the positive and negative polarized photons for the overall process representing  the sum/difference between the quark states' helicities that arises as the quarks emerge from the nucleons. 

The quark helicity or chirality is conserved in this hard subprocess for DVCS. Hence the $A_{\Lambda^\prime,\lambda^\prime;\Lambda,\lambda}$ will be chiral even.
Eq.(\ref{ampg++})
is the Wilson coefficient from Ref.\cite{Ji1} times a kinematical factor 
that will cancel out when multiplied by the soft part as described below.

The convolution in Eq.(\ref{facto}) yields the following decomposition of the transverse photon helicity amplitudes 
\begin{subequations}
\label{f++}
\begin{eqnarray}
f^S_{++} & = & f_{++,++} + f_{-+,-+}     \nonumber \\  & =& g_{++}^S  \otimes (A_{++,++} + A_{-+,-+} )\\
f^A_{++} & =  & f_{++,++} - f_{-+,-+}     \nonumber \\  & =&  g_{++}^A  \otimes (A_{++,++} - A_{-+,-+} )\\
f^S_{+-} & = & f_{++,+-} + f_{-+,--}     \nonumber \\  & =&  g_{++}^S \otimes (A_{-+,++} +  A_{++,-+}) \\  
f^A_{+-} & = & f_{++,+-} - f_{-+,--}      \nonumber \\  & =&  g_{++}^A  \otimes (A_{-+,++} -  A_{++,-+}) 
\end{eqnarray}
\end{subequations}
where, because of parity conservation,
$A_{--,--} = A_{++,++} $, $A_{-+,-+}  = A_{+-,+-}$, $A_{--,+-} = - A^*_{++,-+} $, and $A_{+-,--} = - A^*_{-+,++}$.      

By calculating explicitly the matrix elements  in Eqs.(\ref{CFF}) and (\ref{CFFA}), using the relations below
\begin{subequations}
\begin{eqnarray}
&&\frac{1}{2 \overline{P}^+}  \overline{U}(P^\prime,\Lambda^\prime) \gamma^+ U(P,\Lambda)   =    \frac{\sqrt{1-\zeta}}{1-\zeta/2} \delta_{\Lambda,\Lambda^\prime}  \\
&&\frac{\sqrt{1-\zeta}}{2 \overline{P}^+}  \overline{U}(P^\prime,\Lambda^\prime) \frac{ i \sigma^{+ \mu}}{2M} \Delta_\mu U(P,\Lambda) = 
\nonumber \\ &&
 \frac{-\zeta^2/4}{\left(1-\zeta/2\right)} \delta_{\Lambda,\Lambda^\prime}   +  \frac{- \Lambda \Delta_1 - i \Delta_2}{2M} \delta_{\Lambda, -\Lambda^\prime} \\
&&\frac{1}{2 \overline{P}^+}  \overline{U}(P^\prime,\Lambda^\prime) \gamma^+ \gamma_5 U(P,\Lambda)  =   \Lambda \frac{\sqrt{1-\zeta}}{1-\zeta/2} \delta_{\Lambda,\Lambda^\prime}  \\
&&\frac{\sqrt{1-\zeta}}{2 \overline{P}^+}  \overline{U}(P^\prime,\Lambda^\prime) \gamma_5 \frac{\Delta^+}{2M}  U(P,\Lambda) 
 =  \nonumber \\ & & 
 \frac{\zeta}{2(1-\zeta/2)} \left( \Lambda \zeta \delta_{\Lambda,\Lambda^\prime} + \frac{ \Delta_1 - i\Lambda \Delta_2}{M} \delta_{\Lambda, -\Lambda^\prime}\right).    
 \end{eqnarray}
\end{subequations}
one obtains  the various helicity amplitudes written in terms of the following combinations  of CFFs for the symmetric part,
\begin{subequations}
\label{connect}
\begin{eqnarray}
f^S_{+,+}  & = &  \frac{\sqrt{1-\zeta}}{1-\zeta/2}  {\cal H}  +  \frac{-\zeta^2/4}{\left(1-\zeta/2\right)\sqrt{1-\zeta}} \: {\cal E} \\
f^S_{+,-}   & = & \frac{1}{\sqrt{1-\zeta}}  \frac{1}{1-\zeta/2} \frac{\Delta_1 + i \Delta_2}{2M} \: {\cal E}  
\end{eqnarray}
\end{subequations}
and 
\begin{subequations}
\label{connect2}
\begin{eqnarray}
f^A_{+,+} & =&  \frac{\sqrt{1-\zeta}}{1-\zeta/2}  \widetilde{\cal H}  +  \frac{-\zeta^2/4}{\left(1-\zeta/2\right)\sqrt{1-\zeta}} \: \widetilde{\cal E} \\
f_{+,-}^A & = & \frac{\zeta}{\sqrt{1-\zeta}}  \frac{1}{1-\zeta/2} \frac{\Delta_1 + i \Delta_2}{2M} \: \widetilde{\cal E}  
\end{eqnarray}
\end{subequations}
for the anti-symmetric component.

A similar formalism was presented in \cite{Diehl_rev} where, however, the helicity amplitudes were identified with the combinations,
\[f^S_{++} + f^A_{++}, 
f^S_{++} - f^A_{++}, 
f^S_{+-} + f^A_{+-}, f^S_{+-} - f^A_{+-}, \] 
Here, differently from \cite{Diehl_rev}, we distinguish between the two possible circular polarizations for the transverse virtual photon, 
that generate the $S$ and $A$ components. Such components are written out explicitly throughout this paper.

\subsection{The Hybrid Model}
We evaluate the quark-parton helicity amplitudes in Eq.(\ref{facto}) 
using a covariant model.
The simplest realization of the covariant formalism is the quark-diquark model in which the initial proton dissociates into  a quark
and a recoiling fixed mass system with quantum numbers of a diquark (Fig.1). The covariant model can be made more 
general by letting the mass of the diquark system vary according to a spectral distribution. Extending the diquark mass values generalization 
corresponds to ``reggeizing" the covariant model since the spectral distribution can then reproduce the Regge behavior which is necessary to describe the low $X$ behavior
(a likewise scenario was considered in the pioneering work of  Ref.\cite{BroCloGun}) . Keeping this in mind, in this paper we will include a Regge term multiplicatively, as shown later on.
Because we introduce similarities, or open possible connections with the dual model of Ref.\cite{Pol}, we denote this model of GPDs the {\em hybrid model}. 

For reasons explained in the Introduction, we adopt the diquark model only for the DGLAP region,  where $X \geq \zeta$. The DGLAP region can be considered 
a direct extension of the parton model, where the struck quark with initial longitudinal momentum fraction $X$, is reinserted in 
the proton target after reducing its momentum fraction  to $X-\zeta$, $\zeta$ being the fraction transferred in the exclusive reaction.  
In the DGLAP region the initial and final quarks are both off-shell, while the diquark intermediate state is on mass shell. 
The soft part, $A_{\Lambda^\prime,\lambda^\prime;\Lambda,\lambda}$
is described described in terms of GPDs. $A$ is  given by an integral over the $k^-$ and $k_\perp$ variables (see appendix for detailed expressions).
In the DGLAP region 
the three soft propagators corresponding to the  quark that is emitted $(k^2 -m^2)^{-1}$, the quark that is reabsorbed $(k^{\prime \, 2}-m^2)^{-1}$, and to the intermediate diquark system $(P_X^2-M_X^2)^{-1}$, have poles that lie respectively on the negative imaginary $k^-$ axis ($k$, $k^\prime$), and on the positive axis ($P_X$). Therefore one closes the integration contour on the positive side,  and the diquark is on its mass shell (see also \cite{BroEst}).    

As for the spin structure of the propagators, we have adopted the same scheme as in Refs.\cite{AHLT1,AHLT2} where we considered both the $S=0$ (scalar) and $S=1$ (axial vector) configurations for the diquark. This allows one to obtain distinct predictions for the $u$ and $d$ quarks. However, we assume a similar form 
for the scalar and axial-vector couplings (scalar-like), and we  distinguish their different contributions by varying their respective mass parameters in the calculations.  This assumption is in line with previous estimates \cite{Muld1,AHLT1,AHLT2} where it was advocated that the full account of the axial-vector coupling does not sensibly improve 
the shape of parametrizations, while considerably increasing the algebraic complexity of the various structures (see {\it e.g.} \cite{GGS}). We define $\Gamma$ as the  scalar
coupling at the proton-quark-diquark vertex
\[ \Gamma = g_s \frac{k^2-m^2}{(k^2- M_\Lambda^2)^2}, \] $g_s$ being a constant 
\footnote
{The choice of this vertex function is motivated by phenomenological reasons -- it allows for an easier fit of the form factors -- and also because it agrees with predictions within the Schwinger-Dyson formalism. See discussions in \cite{AHLT1,AHLT2} and in \cite{BroEst}.}.
%
The quark-proton helicity amplitudes are defined as
\begin{equation} 
\label{As}
A_{\Lambda^\prime,\lambda^\prime;\Lambda,\lambda}
  =  \int d^2k_\perp\phi^*_{\lambda^\prime,\Lambda^\prime}(k^\prime,P^\prime) \phi_{\lambda,\Lambda}(k,P),
\end{equation} 
with 
\[\phi_{\lambda,\Lambda}(k,P) =  \Gamma(k) \frac{\bar{u}(k,\lambda) U(P,\Lambda)}{k^2-m^2} \]
\[\phi^*_{\Lambda^\prime \lambda^\prime}(k^\prime,P^\prime) = \Gamma(k^\prime) \frac{\overline{U}(P^\prime,\Lambda^\prime) u(k^\prime,\lambda^\prime)}{k^{\prime \,2}-m^2},  \]
defining the helicity structures at each soft vertex. 
One has: $\phi_{\Lambda \lambda} = \phi^*_{\Lambda \lambda}$, for $\Lambda=\lambda$, $\phi_{\Lambda \lambda} = - \phi^*_{\lambda \Lambda}$
for $\Lambda=-\lambda$.
%
We list the separate structures appearing in Eqs.(\ref{As})
\begin{subequations}
\label{amp3} 
\begin{eqnarray}
A_{++,++} & = & \int d^2k_\perp \phi^*_{++}(k^\prime,P^\prime) \phi_{++}(k,P) \\
A_{+-,+-} & = & \int d^2k_\perp\phi^*_{+-}(k^\prime,P^\prime) \phi_{+-}(k,P)     \\
A_{-+,++} & = & \int d^2k_\perp\phi^*_{-+}(k^\prime,P^\prime) \phi_{++}(k,P)   \\
A_{++,-+} & = & \int d^2k_\perp\phi^*_{++}(k^\prime,P^\prime) \phi_{-+}(k,P).
\end{eqnarray}
\end{subequations}
%

Finally, the denominators are evaluated with the diquark mass on shell,
\begin{subequations}
\label{denom}
\begin{eqnarray}
k^2-m^2 & = & X M^2 - \frac{X}{1-X} M_X^2 - m^2  - \frac{k_\perp^2}{1-X}  \\ 
k^{\prime \, 2}-m^2 & = & \frac{X-\zeta}{1-\zeta} M^2 - \frac{X-\zeta}{1-X} M_X^2 - m^2  - \frac{1-\zeta}{1-X} 
\nonumber \\
& \times & \left( {\bf k}_\perp - \frac{1-X}{1-\zeta} {\bf \Delta}_\perp \right)^2.
\end{eqnarray}
\end{subequations}

To extract the GPDs we calculate the convolutions in Eqs.(\ref{f++}) using the expressions for the $g$ and $A$ functions evaluated above. We obtain
\begin{widetext}
\begin{eqnarray}
& \mathcal{N}  \displaystyle\frac{\sqrt{1-\zeta}}{1-X}  & \int d^2 k_\perp
\frac{%
\left[  \left(m+M X\right)  \left(m + M \frac{X-\zeta}{1-\zeta} \right) + {\bf k}_\perp\cdot \tilde{{\bf k}}_\perp\right]}{(k^2-M_\Lambda^2)^2(k^{\prime \, 2}-M_\Lambda^2)^2} 
 =  \frac{\sqrt{1-\zeta}}{1-\zeta/2}  H  +  \frac{-\zeta^2/4}{\left(1-\zeta/2\right)\sqrt{1-\zeta}} E 
 \\
& \mathcal{N}  \displaystyle\frac{\sqrt{1-\zeta}}{1-X}  &  \int  d^2 k_\perp
 \frac{%
 \left[ \left(m+M X\right)  (\tilde{k}_1 + i \tilde{k}_2)   -  \left(m + M \frac{X-\zeta}{1-\zeta} \right)(k_1 + i k_2) \right]}
 {(k^2-M_\Lambda^2)^2(k^{\prime \, 2}-M_\Lambda^2)^2} 
  =   \frac{1}{\sqrt{1-\zeta}(1-\zeta/2)} \frac{\Delta_1 + i \Delta_2}{2M} E  
\end{eqnarray}

\begin{eqnarray}
&  \mathcal{N} \displaystyle\frac{\sqrt{1-\zeta}}{1-X} &  \int  d^2k_\perp  
\frac{%
\left[  \left(m+M X\right)  \left(m + M \frac{X-\zeta}{1-\zeta} \right) - {\bf k}_\perp\cdot \tilde{{\bf k}}_\perp\right]}{(k^2-M_\Lambda^2)^2(k^{\prime \, 2}-M_\Lambda^2)^2} 
 = \frac{\sqrt{1-\zeta}}{1-\zeta/2}  \widetilde{H}  +  \frac{-\zeta^2/4}{\left(1-\zeta/2\right)\sqrt{1-\zeta}} \widetilde{E} 
 \\
& \mathcal{N}  \displaystyle\frac{\sqrt{1-\zeta}}{1-X}  & \int  d^2k_\perp  \frac{ 
 \left[ \left(m+M X\right)  (\tilde{k}_1 + i \tilde{k}_2)   + \left(m + M \frac{X-\zeta}{1-\zeta} \right)(k_1 + i k_2) \right]}
 {(k^2-M_\Lambda^2)^2(k^{\prime \, 2}-M_\Lambda^2)^2}
 =   \frac{\zeta/2}{\sqrt{1-\zeta}(1-\zeta/2)} \frac{\Delta_1 + i \Delta_2}{2M} \widetilde{E}  
\end{eqnarray}
\end{widetext}
\noindent
from which the following forms for $H$, $E$, $\widetilde{H}$, and $\widetilde{E}$  can be derived
\begin{widetext}
\begin{eqnarray}
H & = &  \displaystyle\mathcal{N} \frac{(1-\zeta)(1-\zeta/2)}{1-X}  \int d^2k_\perp 
\frac{%
\left[  \left(m+M X\right)  \left(m + M \displaystyle\frac{X-\zeta}{1-\zeta} \right) + {\bf k}_\perp\cdot \tilde{{\bf k}}_\perp\right]}{(k^2-M_\Lambda^2)^2(k^{\prime \, 2}-M_\Lambda^2)^2}    +  \frac{\zeta^2}{4(1-\zeta)}E,
\label{GPDH}
\end{eqnarray} 
\begin{eqnarray}
E & = & \displaystyle \mathcal{N} \frac{1}{1-X} \int  d^2k_\perp 
 \frac{-2M (1-\zeta)%
 \left[  \left(m+M X \right) \displaystyle   \frac{\tilde{k} \cdot \Delta}{\Delta_\perp^2}  - \left(m + M \frac{X-\zeta}{1-\zeta} \right) \frac{k_\perp \cdot \Delta}{\Delta_\perp^2} \right] }{(k^2-M_\Lambda^2)^2(k^{\prime \, 2}-M_\Lambda^2)^2}
 \label{GPDE}
\end{eqnarray}

\begin{eqnarray}
\widetilde{H} & = &  \displaystyle\mathcal{N} \frac{1-\zeta/2}{1-X}   \int d^2k_\perp
\frac{%
\left[  \left(m+M X\right)  \left(m + M \displaystyle\frac{X-\zeta}{1-\zeta} \right) - {\bf k}_\perp\cdot \tilde{{\bf k}}_\perp\right]}{(k^2-M_\Lambda^2)^2(k^{\prime \, 2}-M_\Lambda^2)^2}    +  \frac{\zeta^2}{4(1-\zeta)}\widetilde{E}
\label{GPDHTILDE}
\end{eqnarray} 
\begin{eqnarray}
\widetilde{E} & = & \displaystyle\mathcal{N} \frac{1-\zeta/2}{1-X} \int d^2k_\perp
\frac{- \displaystyle\frac{4M (1-\zeta) }{\zeta}%
 \left[  \left(m+M X \right) \displaystyle   \frac{\tilde{k} \cdot \Delta}{\Delta_\perp^2}  + \left(m + M \frac{X-\zeta}{1-\zeta} \right) \frac{k_\perp \cdot \Delta}{\Delta_\perp^2} \right] }{(k^2-M_\Lambda^2)^2(k^{\prime \, 2}-M_\Lambda^2)^2}
 \label{GPDETILDE}
\end{eqnarray}
\end{widetext}
where $\mathcal{N}$ is in GeV$^4$.
The integrations over $d^2 k_\perp$ yield finite values for the amplitudes in the limit $\Delta \rightarrow 0$ and $\zeta \rightarrow 0$ (see Appendix \ref{appA}).

\subsection{Reggeization}
\label{reggeiz}
It was noticed in Refs.\cite{AHLT1,AHLT2} that the low $X$ behavior of the GPDs in the forward limit necessitates an extra factor of the type $\approx X^{-\alpha}$ in order to adequately fit current DIS data. We reiterate that, for the off-forward case this factor is important even at intermediate/large values of $X$ and $\zeta$ because the dominant behavior of the GPDs at low $X$ determines the nucleon form factors. In other words the lack of such a term hinders a good fit of the form factors.

The Regge term indeed can be seen as originating from a generalization of the diquark picture in which 
the mass of the diquark, $M_X^2$ is not fixed, but has a spectral distribution. As first shown in Ref.~\cite{BroCloGun} in a simple covariant model for a pdf, choosing a spectral distribution of the form $ \rho_R(M_X^2) \propto M_X^{2\alpha(0)}$ gives rise to the pdf behavior $x^{-\alpha(0)}$ upon integration over all $M_X^2$. A more physical picture of the nucleon might have two contributions,
\begin{equation}
\rho(M_X^2) = \rho_G(M_X^2) + \rho_R(M_X^2),
\label{rho_regge}
\end{equation}
wherein $\rho_G$ is sharply peaked at a fixed value of the diquark mass $M_X^2= \overline{M}_X^2$, and it yields the usual "fixed mass" diquark model.  $\rho_R$ smoothly extends $M_X^2$ to large values and, with the Regge power behavior, upon integration over $M_X^2$, will  yield a Regge type behavior largely dictated by the shape of the spectral function itself . 

\begin{figure}
\includegraphics[width=9.cm]{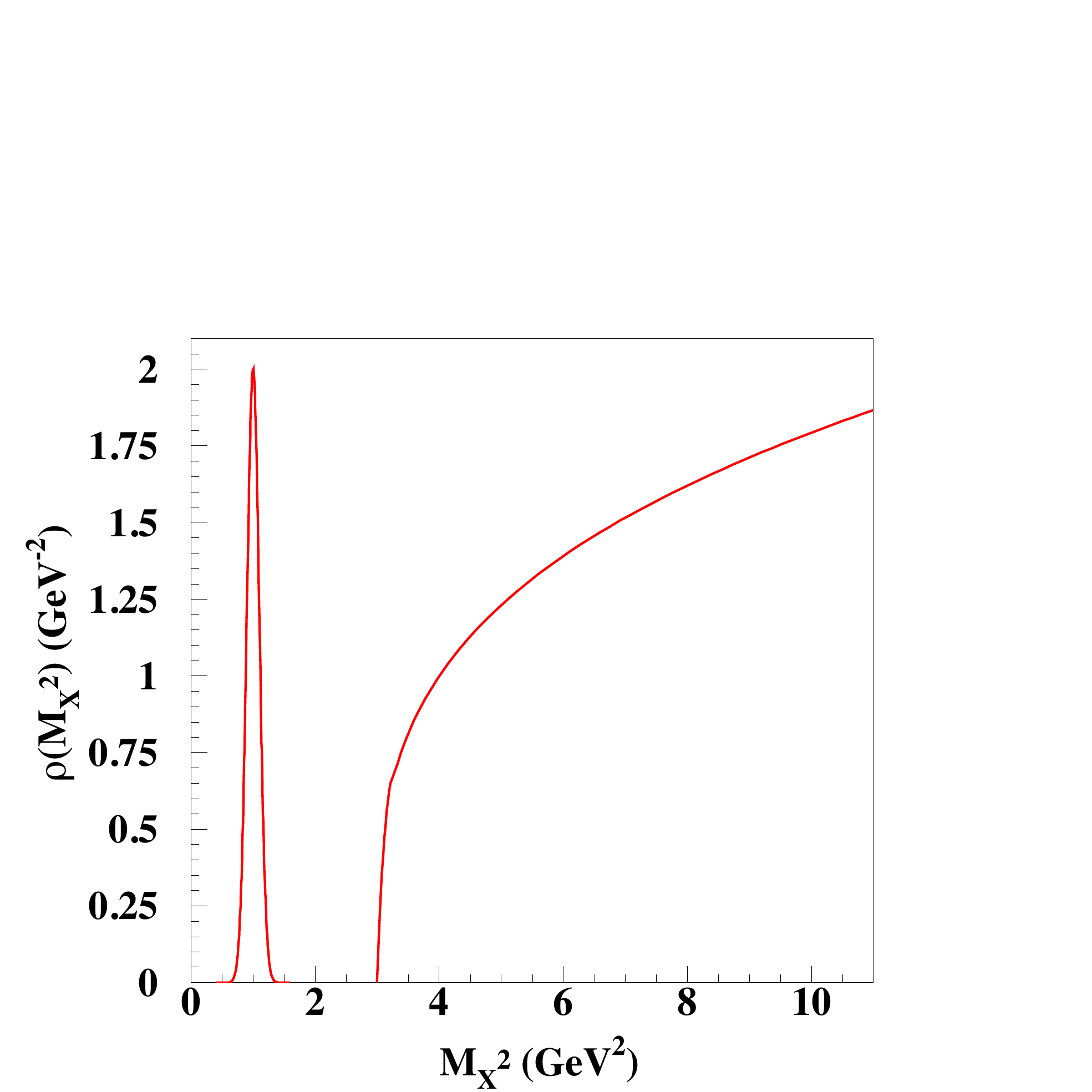}
\caption{A plausible diquark mass distribution, the spectral density with Regge behavior $M_X^{2\alpha(0)}$ with $\alpha(0)\approx \frac{1}{3}$.}
\label{figrho}
\end{figure}

We will illustrate this "reggeization" process by considering the spin independent GPD $H(X,0,0)=f_1(X)$ as a function of a continuum of diquark masses. Aside from overall constant factors, the exact expression for the forward limit for the mass $M_X$, obtained from Eq.(~\ref{GPDH}),  is given by
 \begin{eqnarray}
 H(X,0,0)&\propto&
\left[\frac{2(m_q+XM)^2}{[M_X^2+\frac{(1-X)}{X}(M_\Lambda^2-XM^2)]^3} \right. \\
 \nonumber
 & & \left. + \frac{X}{[M_X^2+\frac{(1-X)}{X}(M_\Lambda^2-XM^2)]^2}\right] \frac{\pi (1-X)^4}{6X^3}
 \label{HX00}
\end{eqnarray}
Multiplying this expression with $\rho_R(M_X^2)\propto M_X^{2\alpha(0)}$ and then integrating over all diquark masses from 0 to $\infty$ gives the analytic result 
\begin{equation}
\int_0^\infty dM_X^2 \rho_R(M_X^2) H(X,0,0)  \sim X^{-\alpha(0) -1},
\end{equation}
for $X\rightarrow 0$. In practice we would not integrate from zero mass and we would cut off the integral at the maximum mass allowed by the kinematics. The fixed mass term $\rho_G(M_X^2)$ would give back the unintegrated expression Eq.(\ref{HX00}) which goes to a constant or $X^0$ as $X\rightarrow 0$. A plausible form for the spectral density of the diquark is shown in Figure~\ref{figrho}.

Our main conclusion is  that we can generate the Regge behavior and still be consistent with the diquark model. 

At this point this discussion does not consider the $t$-dependence, although it is plausible to incorporate the Regge trajectory form $\alpha(t)=\alpha_0 + \alpha^\prime t$, while including the $t$-dependence of the diquark model. Including the skewedness, $\zeta \neq 0$ is more complicated, because of the distinction between the DGLAP and ERBL regions. 
While detailed calculations using Eq.(\ref{rho_regge}) will be presented elsewhere,
in the present analysis we adopt
a factorized form of the Regge term which has  similar features as the Reggeized diquark model, but is more in line
with current parametric forms of parton distributions,  
\begin{equation}
R =  X^{-[\alpha + \alpha^\prime(X) t + \beta(\zeta) t ]},
\label{regge}
\end{equation}
In this form only the term $X^{-\alpha}$ can be considered a proper Regge contribution. 
The term $ \alpha^\prime(X) =  \alpha^\prime  (1-X)^{p} $ is constructed so as to guarantee that upon Fourier transformation in ${\bf \Delta}_\perp$, one obtains finite values for the nucleon radius as $X\rightarrow 1$ \cite{Bur_radius}. This prescription is obtained for $\zeta=0$, or $t \equiv - \Delta_\perp^2$ \cite{gpdshort}. In order to extend it to $\zeta \neq 0$, an additional term in the exponent, $\beta(\zeta)$,  is introduced.  The physical motivation for  this term 
is that it effectively accounts for the shift between the initial and final proton's coordinates that occurs when Fourier transforming GPDs at $\zeta \neq 0$ 
\cite{Diehl_radius}. 

As we will show in Sections \ref{sec3} and \ref{sec4}, two forms of $\beta$ are suggested by the behavior of the available DVCS data, 
\begin{eqnarray}
\label{betaI}
\beta_I(\zeta) & = & \beta \frac{\zeta^2}{1-\zeta} \\
\label{betaII}
\beta_{II}(\zeta) & = &  \beta \zeta^a.
\end{eqnarray}
The effect of these terms is to allow for a data driven change in the slope in $\zeta$ of the GPDs, most likely an increase,  with respect to the diquark model predictions. 
Since most DVCS data so far appear as asymmetries given by ratios of different cross sections combinations, it is difficult to determine precisely the $\zeta$ behavior of the CFFs and GPDs. It is therefore indispensable in future experimental analyses to provide absolute cross sections, as already done for the set of data provided by Hall A.      
Introducing directly DVCS data to determine the behavior of our fit is an important step that distinguishes our analysis from other ones in that it helps establishing the main trends of the multi-variable dependent data. The treatment of multi-variable dependence characterizes analyses aimed at extracting GPDs from data. What we suggest here is a bottom-up approach where experimental evidence is used to constrain the various theoretical aspects of the GPDs behavior, eventually giving rise to a complete model.

\subsection{Crossing Symmetries}
GPDs observe precise crossing symmetry relations.
In order to discuss these symmetry properties we first introduce the so-called "symmetric system"
of variables $\{x,\xi \}$, where 
\[ x= \frac{k^+ + k^{\prime \, +}}{P^+ + P^{\prime \, +} }= \frac{X-\zeta/2}{1-\zeta/2} \]
\[ \xi = \frac{2 \Delta^+}{P^+ + P^{\prime \, +} } = \frac{\zeta}{2 - \zeta} \] 

We also introduce the quark labels for all four chiral even GPDs, $F_q \equiv \{H_q, E_q, \widetilde{H}_q, \widetilde{E}_q \}$. 
By analogy with DIS, 
we define $F_q(x,\xi)$ in the interval $-1 \leq x \leq 1$, with the following identification of anti-quarks,
\begin{eqnarray}
F_{\bar{q}}(x,\xi) & = & -  F_q(x,\xi)    \; \; \;  x<0.                            
\end{eqnarray}
From this expression one defines 
\begin{eqnarray}
F_q^- & = & F_q(x,\xi) - F_{\bar{q}}(x,\xi) \\
F_q^+ & = & F_q(x,\xi) +  F_{\bar{q}}(x,\xi), 
\end{eqnarray} 
where $F_q^-$ is identified with the  flavor non singlet,  valence quarks distributions, and $\sum_{q} F_q^+$ with the flavor singlet, sea quarks distributions. $F_q^-$ and $F_q^+$ obey the symmetry relations
\begin{eqnarray}
F_q^-(x,\xi)  & = & F_q^-(-x,\xi)  \\
F_q^+(x,\xi)  & = & - F_q^+(-x,\xi). 
\end{eqnarray} 
In DIS 
the commonly adopted Kuti-Weisskopf model ensues \cite{Kuti} by which all distributions are evaluated at positive $x$. 

In the off-forward case crossing symmetries are important for the evaluation of the CFFs defined in Eq.(\ref{direct}).  The Wilson coefficient function, in fact, also obeys crossing symmetry relations,
\begin{eqnarray}
C^\pm(x,\xi) =  \frac{1}{x-\xi + i \epsilon} \pm  \frac{1}{x+\xi - i \epsilon},  
\end{eqnarray}
so that 
\begin{eqnarray}
\mathcal{H}_q & = &  \int_0^1 dx  \, C^+(x,\xi)  H_q^+(x,\xi,t)\\
\widetilde{\mathcal{H}}_q & = & \int_0^1 dx  \, C^-(x,\xi)  \widetilde{H}_q^-(x,\xi,t),
\end{eqnarray}
similar relations hold for $E_q$ and $\widetilde{E}_q$.
In the non-symmetric system of variables adopted throughout this paper,  the axis of symmetry is shifted to $X=\zeta/2$. Moreover,  $x  \in [-1,1] \Rightarrow X \in [-1+\zeta, 1]$ , and $x=-\xi \Rightarrow X=0$, $x=\xi \Rightarrow X=\zeta$.  
\begin{figure}
\includegraphics[width=9.cm]{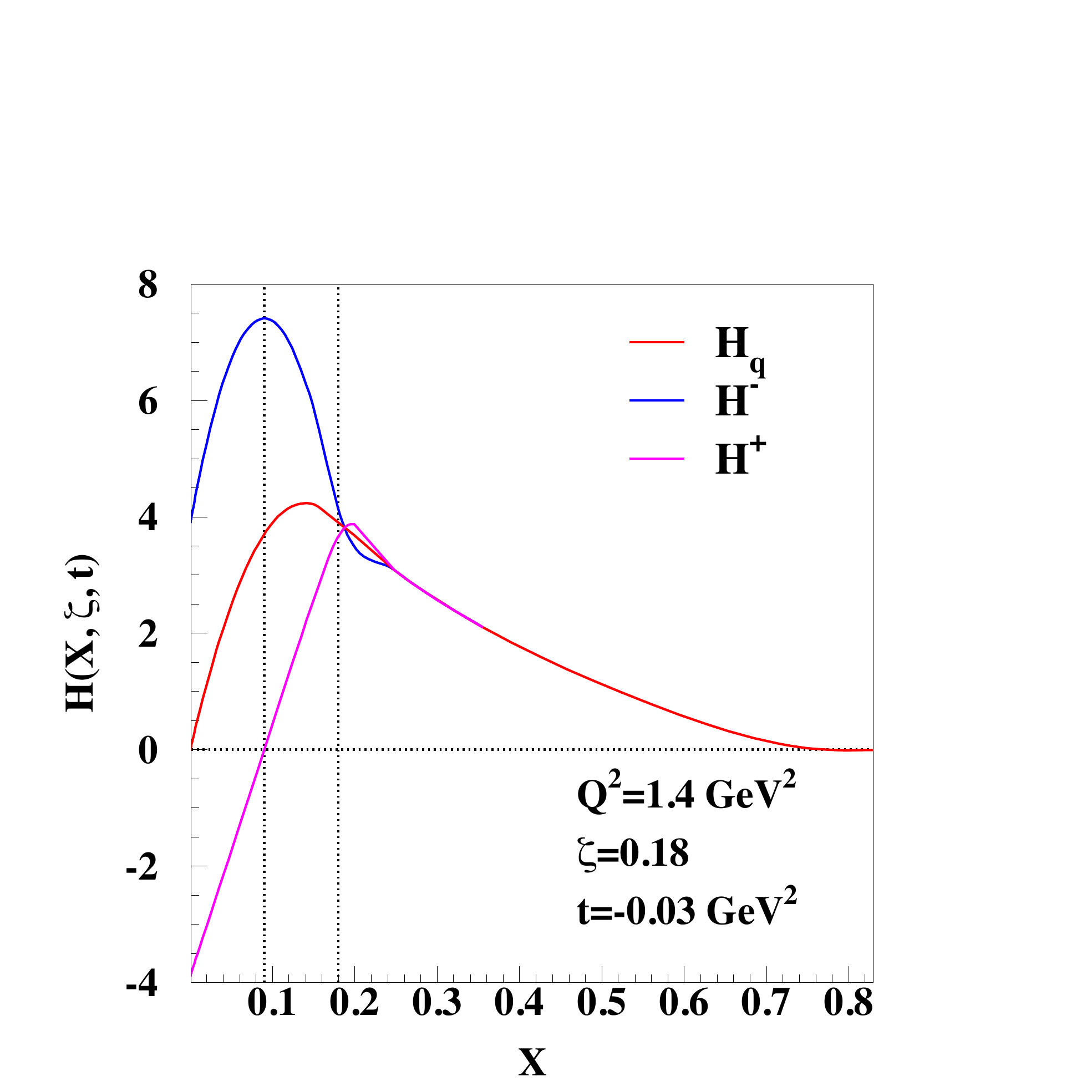}
\caption{(color online) Crossing symmetric, $H^+$, and antisymmetric, $H^-$, contributions to the GPD $H^q$ at $\zeta=0.18$, and $t=t_{min}$.
In this example, $H^{\bar{q}}$ was taken equal to zero in the DGLAP region.}
\label{fig4}
\end{figure}

As explained in the Introduction, the validity of a simple handbag based partonic interpretation of DVCS in the ERBL region has been recently questioned. 
The safest choice for a parametrization in the ERBL region is therefore to adopt a "minimal" model that accounts for crossing symmetry properties, 
continuity at the crossover points ($X=0$ and $X=\zeta$), and polynomiality.
A possible form is  obtained by parametrizing the crossing symmetric and anti-symmetric contributions, respectively as follows
\begin{eqnarray}
\label{h+}
H^-_{X<\zeta}(X,\zeta) & = & a^-(\zeta) X^2 -a^-(\zeta) \zeta X + H(\zeta,\zeta) \\
\label{h-}
H^+_{X<\zeta}(X,\zeta) & = & a^+ X^3 -a^+ \zeta X^2 + c X+ d   
\end{eqnarray}
where $a^-(\zeta) = 6(\zeta H(\zeta,\zeta) - 2 S_{ERBL})/\zeta^3$, $S_{ERBL}$ being the area subtended by $H^-_q$ in the ERBL region, 
\begin{eqnarray}
S_{ERBL} & = & \int_0^\zeta  dX \, H^-_q(X,\zeta,t) = \nonumber \\
&=& \left(1-\frac{\zeta}{2} \right) \left(F_1^q - \int_\zeta^1 \frac{H^q(X,\zeta,t)}{1-\zeta/2} dX \right).
\end{eqnarray}
$S_{ERBL}$ appears in the definition of $a^-$ multiplied by a factor of $2$ because of the crossing symmetry property for the
areas subtended by $H^-_q$ and $H_q$. 
In Eq.(\ref{h-}) $a^+$ is a free parameter. This choice of parameters gives $H_q=H^-=H^+$ at $X=\zeta$, the antiquark component in the DGLAP region 
being taken to be equal to zero in this phase of our analysis. 
Notice that $H_q$ and $H_{\bar{q}}$ are not required to obey crossing symmetries. They are obtained by construction from Eqs.(\ref{h+},\ref{h-}). 
An example describing the symmetric and antisymmetric components of  $H^q$ is given in Fig.\ref{fig4}.  

  
\section{Recursive fit: Numerical evaluation of GPD parameters from inclusive measurements constraints.}
\label{sec3}
Here we describe a recursive fitting procedure to extract the chiral-even GPDs from available DVCS data.
Our fit uses the parametric forms (we omit the quark labels for simplicity)
\begin{equation}
F(X,\zeta,t)  = {\cal N} G_{M_X,m}^{M_\Lambda}(X,\zeta,t) \,  
R^{\alpha,\alpha^\prime}_p(X,\zeta,t) 
%
\label{fit_form}
\end{equation}
where $F \equiv H,E,\widetilde{H},\widetilde{E}$; the functions $G_{M_X,m}^{M_\Lambda}$ are the covariant diquark contributions from Eqs.(\ref{GPDH},\ref{GPDE},\ref{GPDHTILDE},\ref{GPDETILDE}), and $R^{\alpha,\alpha^\prime}_p$ was given in Section \ref{reggeiz}.
\footnote{
In Appendix \ref{appB} we present additional parametric forms that are more practical for applications and numerical calculations. }
  
It should be remarked that our new parametric form follows from the one used in Ref.\cite{AHLT1,AHLT2}, while presenting several important differences. 
We have first of all, completed a thorough analysis of the spin components of the various GPDs, both in the unpolarized and polarized sectors, thus releasing the assumption of a simplified quark-proton vertex structure made in \cite{AHLT1,AHLT2}, and extending our analysis to the much needed $\widetilde{H}$ and $\widetilde{E}$ functions. The more careful spin treatment also results in a different shape for $E$ which in \cite{AHLT1,AHLT2} closely followed $H$ by construction.

The most important features of our new parametrization are summarized below:

\noindent 
{\it i)} we consider only configurations for a spin $1/2$ quark and a spin $0$ diquark. The flexibility
in shape contributed by considering a spin $1$ diquark was in fact not sufficient to allow us to model
{\it e.g.} the rise at low $X$. We therefore opted for keeping the Regge term as in \cite{AHLT1,AHLT2}.
This can in fact be derived from a "reggeized" version of the model  as explained in Section \ref{sec2}.

 \noindent 
{\it ii)} we model {\em all} chiral-even GPDs, and we present for the first time parametric forms for $\widetilde{H}$ and $\widetilde{E}$, besides new evaluations for $H$ and $E$. Our analysis applies
to the intermediate $x_{Bj}$, multi-GeV $Q^2$ regime which is dominated by valence quarks in the DGLAP region. Only $u$ and $d$  quark flavors are considered. 

\noindent 
{\it iii)} we perform a recursive fit in which parameters are evaluated orderly,  from imposing constraints from DIS experimental results first, then from the elastic form factors, and eventually including DVCS data directly.  This procedure affords us a better control on: {\it i)} the number of parameters that are necessary to constrain the GPD multi-variable problem; {\it ii)} the fit's variants as new data are inserted.   

All parameters obtained from the DIS and elastic constraints are given in Table \ref{table_DIS}. They correspond to the first two steps of our fitting procedure.
While we address in detail the impact of the GPDs $H, E, \widetilde{H}$ on the description of available DVCS data, we  postpone the discussion of  $\widetilde{E}$  to a dedicated
analysis in \cite{GGL_Etilde}. As we explain in what follows $\widetilde{E}$ contributes   
to DVCS observables  multiplied by a factor $x_{Bj}$, or $\xi \approx x_{Bj}/(2-x_{Bj})$.
As experimental data on DVCS target asymmetries accumulate, it is important to clarify that the pion pole
contribution to $\widetilde{E}$ scales as $1/x_{Bj}$ only in specific models like the chiral soliton based factorized form  described in Ref.\cite{Goeke_rev}.  While this factorized form is a convenient model, the $\xi$ singularity  is not required by the general analytic structure of the GPD. In fact, in our evaluation $\widetilde{E}$ is estimated to be suppressed by a factor $ \lesssim 0.1$ at HERMES kinematics. 
\vspace{0.3cm}

The first set of experimental constraints is given by the valence contribution to the 
inclusive DIS structure functions,
\begin{subequations}
\begin{eqnarray}
H^q(X,0,0,Q^2) & = & f_1^q(X,Q^2)  \equiv q_v(X) \\
\widetilde{H}^q(X,0,0,Q^2) & = & g_1^q(X,Q^2) \equiv \Delta q_v(X) 
\end{eqnarray}
\end{subequations}
representing the forward limit of Eq.(\ref{fit_form}) (we have restored both the quark labels $q=u,d$, and for the $Q^2$ dependence).
$f_1^q$ and $g_1^q$ are obtained from DIS data.  We do not use directly experimental data in this phase of the analysis, but we perform a fit of the valence components of existing parametrizations.  The fit was performed similarly to Refs.\cite{AHLT1,AHLT2}.  By inspecting Eqs.(\ref{GPDH}), (\ref{GPDHTILDE}) and (\ref{fit_form}) one can see that 
for $t=0$ and $\zeta =0$, the only parameters that enter are: $M_X, M_\lambda, m$ in $G_{M_X,m}^{M_\Lambda}$, and $\alpha$ in  $R^{\alpha,\alpha^\prime}$. 

\begin{figure}
\includegraphics[width=5.cm]{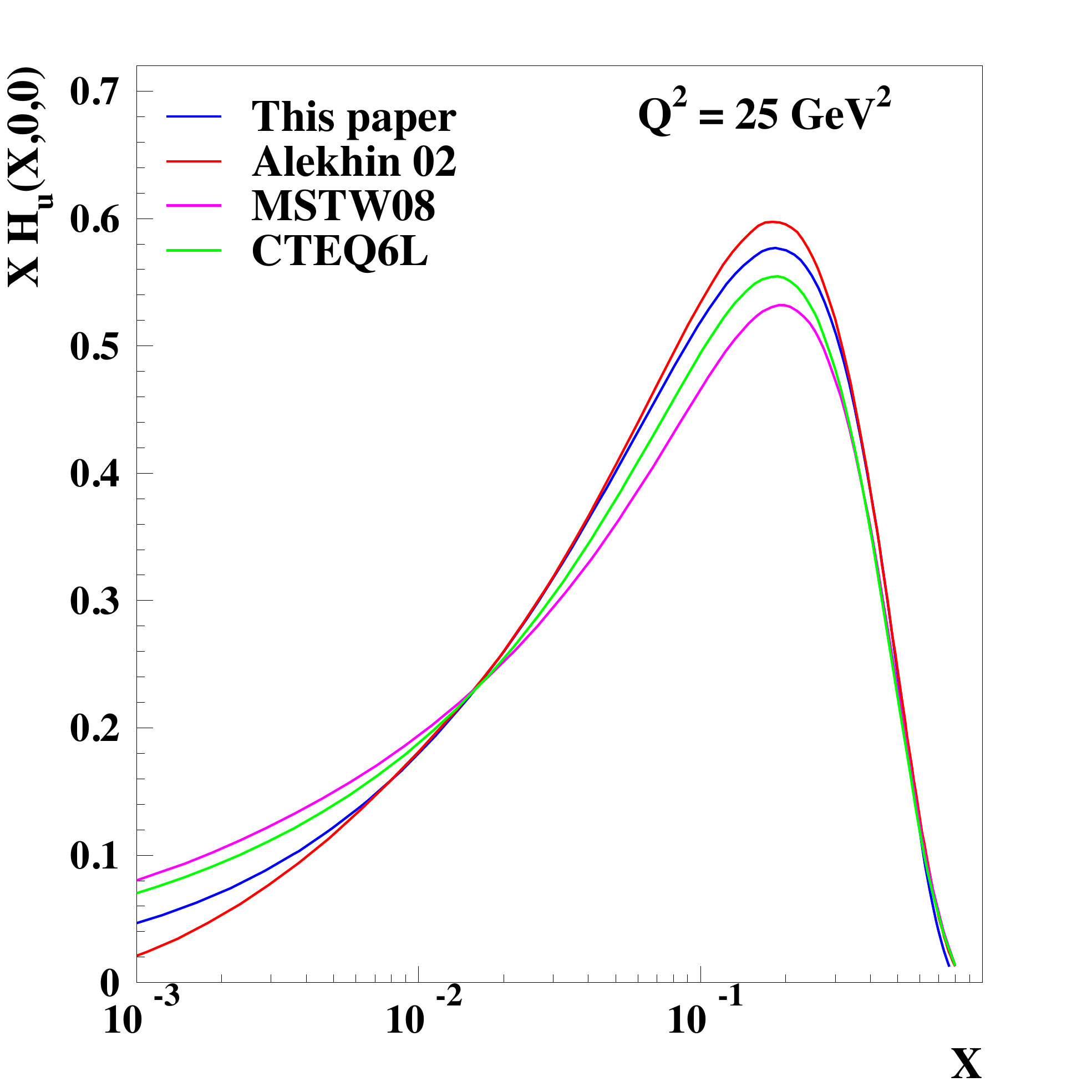}
\includegraphics[width=5.cm]{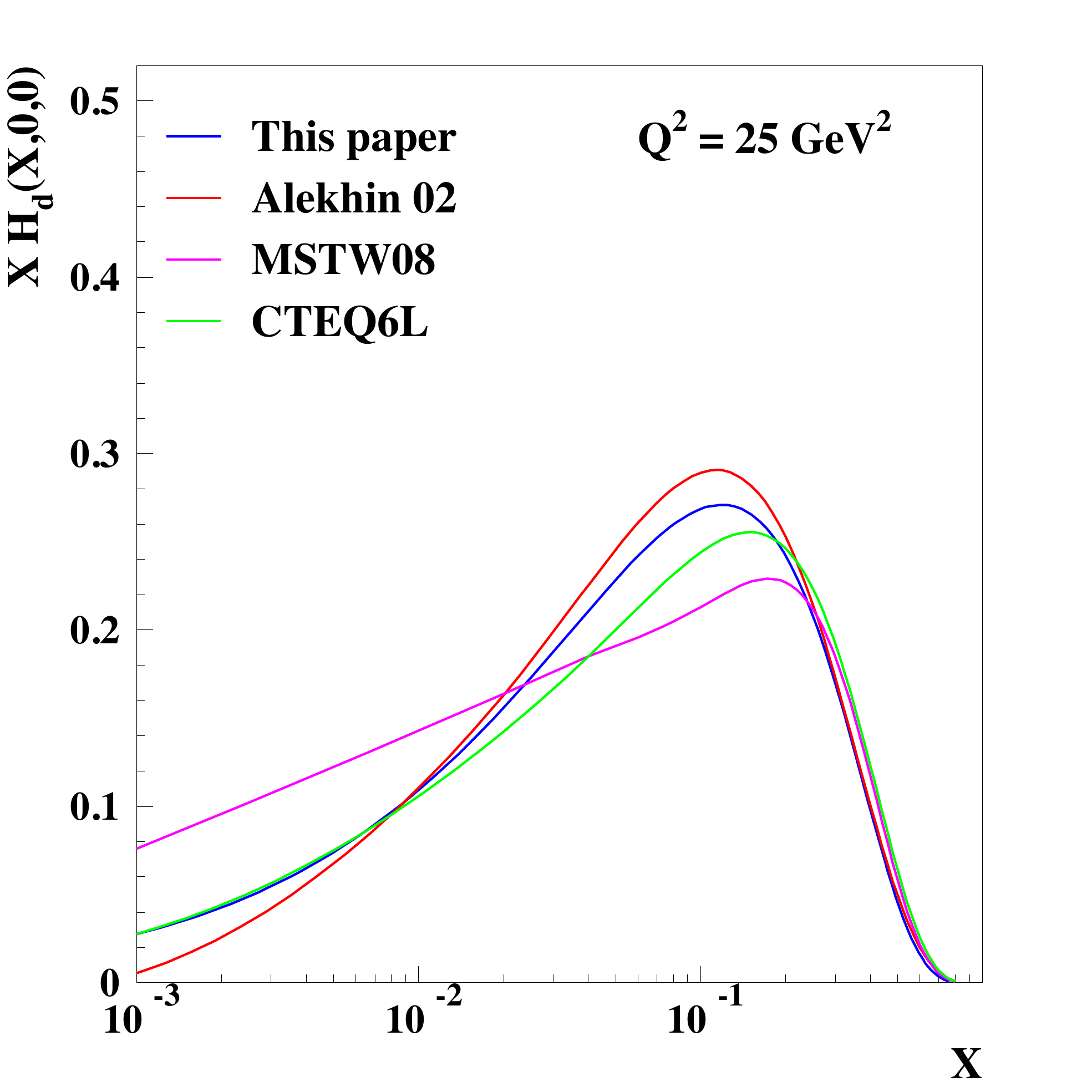}
\caption{(color online) GPDs $H_u(X,0,0)$ (top) and $H_d(X,0,0)$ (bottom), evaluated at $Q^2 =$ 25 GeV$^2$, compared with current LO parametrizations 
\cite{AMP06,MSTW08,CTEQ6.6}.  
\label{fig5} }
\end{figure}

We fit separately the unpolarized, $H$, and polarized, $\widetilde{H}$ GPDs. 
For $H$, an additional parameter, ${\cal N}$ is fixed by the baryon number sum rules constraints, $\int _0^1 dX u_v(X) = 2$, and   $\int_0^1 dX d_v(X) = 1$.
Therefore in our first step we have four parameters per distribution, per quark flavor giving a total number of parameters consistent with what obtained in recent PDF parametrizations {\it e.g.} 
\cite{AMP06,CTEQ6.6,MSTW08}. As already noticed in \cite{AHLT1,AHLT2,Muld1}, the diquark model based parametrization corresponds to a low initial scale, $Q_o^2$.  Parametric forms are then evolved to the $Q^2$ of the data using LO Perturbative QCD (PQCD) evolution equations \cite{Vinnikov}. 
Additional parameters not shown in the Table are the initial value of the perturbative evolution scale, $Q_o^2 = 0.0936$ GeV$^2$, and the parameters $\beta$ appearing
in Eq.(\ref{betaI}), $\beta=10$,  and Eq.(\ref{betaII}), $\beta=1.5$. These were fixed by implementing directly DVCS data in our fit, as we will show in Section \ref{sec4}. 
%
\begin{widetext}
\begin{center}
\begin{table}[h]
\center
\begin{tabular}{|c|c|c|c|c|}
\hline
\hline

Parameters             &  $H$                &  $E$                & $\widetilde{H}$       & $\widetilde{E} $  \\ 
\hline
\hline
$m_u $ (GeV)           & 0.367               &  0.367              &  2.479                &  2.479            \\
$M_X^u$ (GeV)          & 0.583               &  0.583              &  0.467                &  0.467            \\
$M_\Lambda^u$ (GeV)    & 0.963               &  0.963              &  0.909                &  0.909            \\
$\alpha_u$             & 0.222               &  0.222              &  0.218                &  0.218            \\
$\alpha^\prime_u$      & 2.443  $\pm$ 0.063  &  4.582 $\pm$ 0.128  &  1.758 $\pm$ 0.839    &  5.549 $\pm$ 0.519\\
$p_u$                  & 0.6649 $\pm$ 0.0268 &  1.465 $\pm$ 0.031  &  0.558 $\pm$ 0.468    &  0.420 $\pm$ 0.069\\
${\cal N}_u$           & 1.468               &  1.468              &  0.0343 $\pm$ 0.0033  &  4.882 $\pm$ 0.636\\ \hline

$m_d $ (GeV)           & 0.0850              & 0.0850              &  1.211                &  1.211                 \\
$M_X^d$ (GeV)          & 0.841               & 0.841               &  0.699                &  0.699                 \\
$M_\Lambda^d$ (GeV)    & 0.7592              & 0.7592              &  0.836                &  0.836                 \\
$\alpha_d$             & 0.0378              & 0.0378              &  0.0417               &  0.0417                \\
$\alpha^\prime_d$      & 1.777  $\pm$ 0.021  & 0.0516 $\pm$ 0.0026 &  1.489 $\pm$ 0.629    &  4.791  $\pm$ 0.316    \\
$p_d$                  & 0.114  $\pm$ 0.015  & -10.147$\pm$ 0.681  &  1.032 $\pm$ 0.552    &  0.248  $\pm$ 0.060    \\
${\cal N}_d$           & 1.023               & -2.368 $\pm$ 0.160  & -0.0768 $\pm$ 0.0068  & -17.414 $\pm$ 1.815    \\ \hline

\hline
\hline
\end{tabular}
\caption{\label{table_DIS} Parameters obtained from our recursive fitting procedure applied to $H_q$, $E_q$, and $\widetilde{H}_q$, $q=u,d$.  $m_q$, $M_X^q$, $M_\Lambda^q$, and $\alpha_q$ were obtained in a first phase 
by fitting valence quarks PDFs from DIS experimental data.  $\alpha^\prime_q$ and $p_q$ were obtained subsequently, by fitting the proton and neutron elastic form factors -- $H_q$ and $E_q$ --  and the axial form factors -- 
$\widetilde{H}_q$ The value of the additional parameter $\beta$ in Eq.(\ref{regge}) is $\beta=10$.}
\end{table}
\end{center}
\end{widetext}
In the fit for $\widetilde{H}$ we use a similar scheme as in current fits 
\cite{Goto,LSS},  where
\[ X \Delta  q_v(X,Q^2) = {\cal N}_q \, X^{-a_q} Xq_v(X,Q^2). \]  
We left the mass parameters $M_X$ and  $M_\Lambda$ fixed as for the unpolarized case,
while we varied $\alpha$ and $m$. By letting the latter vary, we obtain the effect of the extra term
$ \propto (1 + \gamma X)$ introduced in \cite{LSS} for the LO fit. 
Figure \ref{fig5} shows our curves for  $H_{u,d}(X,0,0)$ at the rather high value of $Q^2=$ 25 GeV$^2$, in order to test the stability of our
fit with PQCD evolution. Other available PDF parametrizations from quantitative fits are also shown in the figure.

In the second phase of our fit we impose an additional set of independent experimental constraints from the normalizations of the chiral even GPDs to the nucleon 
form factors, 
\begin{subequations}
\begin{eqnarray}
\int_0^1 H^q(X,\zeta,t) & = F_1^q(t) \\
\int_0^1 E^q(X,\zeta,t) & = F_2^q(t) \\
\int_0^1 \widetilde{H}^q(X,\zeta,t) & = G_A^q(t) \\ 
\int_0^1 \widetilde{E}^q(X,\zeta,t) & = G_P^q(t) 
\label{GP}
\end{eqnarray}
\end{subequations}
where $F_1^q(t)$ and $F_2^q(t)$ are the Dirac and Pauli form factors for the quark $q$ components in the nucleon. $G_A^q(t)$ and $G_P^q(t)$ are the axial  and pseudoscalar form factors. 

Notice that the GPD $\widetilde{E}$ is constrained by 
the pseudoscalar form factor of the nucleon through Eq.(\ref{GP}).
When the covariant or light front diquark spectator model is applied to calculating  $\widetilde{E}(X,\zeta,t)$, there is no kinematical singularity. The combination $A_{++,-+}+A_{-+,++}$ vanishes as $\zeta \rightarrow 0$ for any $X$ and $t$. This appears as a restriction on the GPD in the DGLAP region, $X \ge \zeta$. Requiring polynomiality leads to the sum rule in the ERBL region
\begin{equation}
\int_0^\zeta \frac{dX}{1-\zeta/2} \widetilde{E}(X,\zeta,t) = G_P(t)-\int_{\zeta}^{+1} \frac{dX}{1-\zeta/2} \widetilde{E}(X,\zeta,t)
\label{FpERBL}
\end{equation}
Since the integral in the DGLAP region is finite for any $\zeta$ and does not have a pole at $\zeta=0$, the ERBL region integral will not either. This will be true of any spectator model wherein there are no kinematic singularities introduced. 
In the diquark spectator approach that we use, the $t$ dependence of the pion pole in the form factor can be reproduced while satisfying the sum rule in Eq.~\ref{FpERBL} (for small $|t|$) by a suitable choice of mass and ``Regge'' parameters. This corresponds to a dual picture - a $t$-channel pion pole emerging from an integral over an $s$-channel diquark pole. 
In summary, we reiterate that  the GPD $\widetilde{E}$ enters the target asymmetry always multiplied by $\zeta$($x_{Bj}$), so that it contributes only weakly in the HERMES kinematical region. 

Isospin decompositions allow one to relate the quark form factors to experimental measurements
of $F_{1(2)}^{p(n)}$, the Dirac (Pauli) form factors for the proton (neutron), respectively. 
$G_{A}$ and $G^o_{A}$ are the isovector and isoscalar components of the axial  nucleon form factor, and $G_P$.   
We used the same selection of data as in Ref.\cite{AHLT1} for the nucleon electromagnetic form factors (see references in \cite{Kelly}).  The resulting parameters are given in Table \ref{table_DIS}. More recent data \cite{Puckett} are now available that show a milder slope of the electric to magnetic proton form factors ratio at large $|t|$. However, these do not largely affect our fits that are limited 
to the $-t << Q^2 \approx 2-3$ GeV$^2$ region.
$G_{A}$ is obtained from the global average of neutron beta decay, and neutrino scattering experiments (see Ref. \cite{Schindler} and references therein),
\begin{equation}
G_A(t) = \displaystyle \frac{g_A}{\left( 1 - \displaystyle \frac{t}{M_A^2} \right)^2}
\end{equation}
with $g_A = 1.2695 \pm 0.0029$, and $M_A = 1.026 \pm 0.021$ GeV.  $G_P$ is notoriously dominated by a pion pole contribution, a small non-pion pole component
being also present. We used the experimental values displayed {\it e.g.} in Ref.\cite{Fearing}. A more thorough discussion of this form factor will be given in  \cite{GGL_Etilde}.

As a result, for each quark flavor and GPD type, using the constraints above, we can determine the additional parameters, $\alpha^\prime$, $p$ in Eq.(\ref{fit_form})
and the normalizations ${\cal N}$ (Eq.(\ref{fit_form}) and Table \ref{table_DIS}).   
The number of parameters used is consistent with the one used in fits of the nucleon form factors data. These require four parameters for $G_E^p$, $G_M^p$, $G_M^n$, respectively, and two for $G_E^n$ \cite{Kelly,Bodek}. 

\begin{figure}
\includegraphics[width=10.cm]{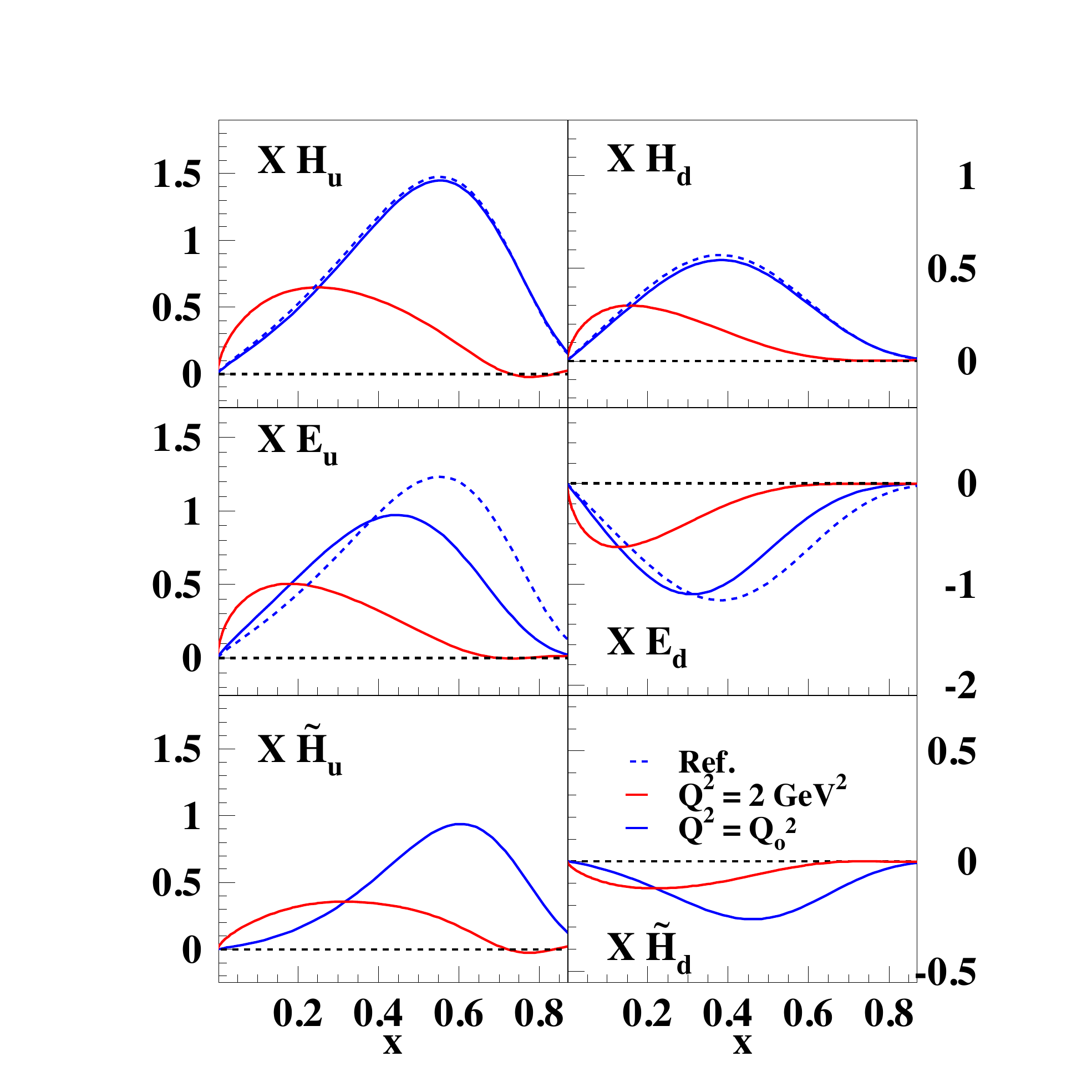}
\caption{(color online) GPDs $F_q(X,0,0) \equiv \{H_q, E_q, \widetilde{H}_q \}$, for $q=u$ (left) and $q=d$ (right), evaluated at the initial scale, $Q_o^2 =$ 0.0936 GeV$^2$, and at $Q^2 =$ 2 GeV$^2$, respectively.  The dashed lines were calculated using the model in Refs.\cite{AHLT1,AHLT2} at the initial scale. 
\label{fig6} }
\end{figure}
\begin{figure}
\includegraphics[width=10.cm]{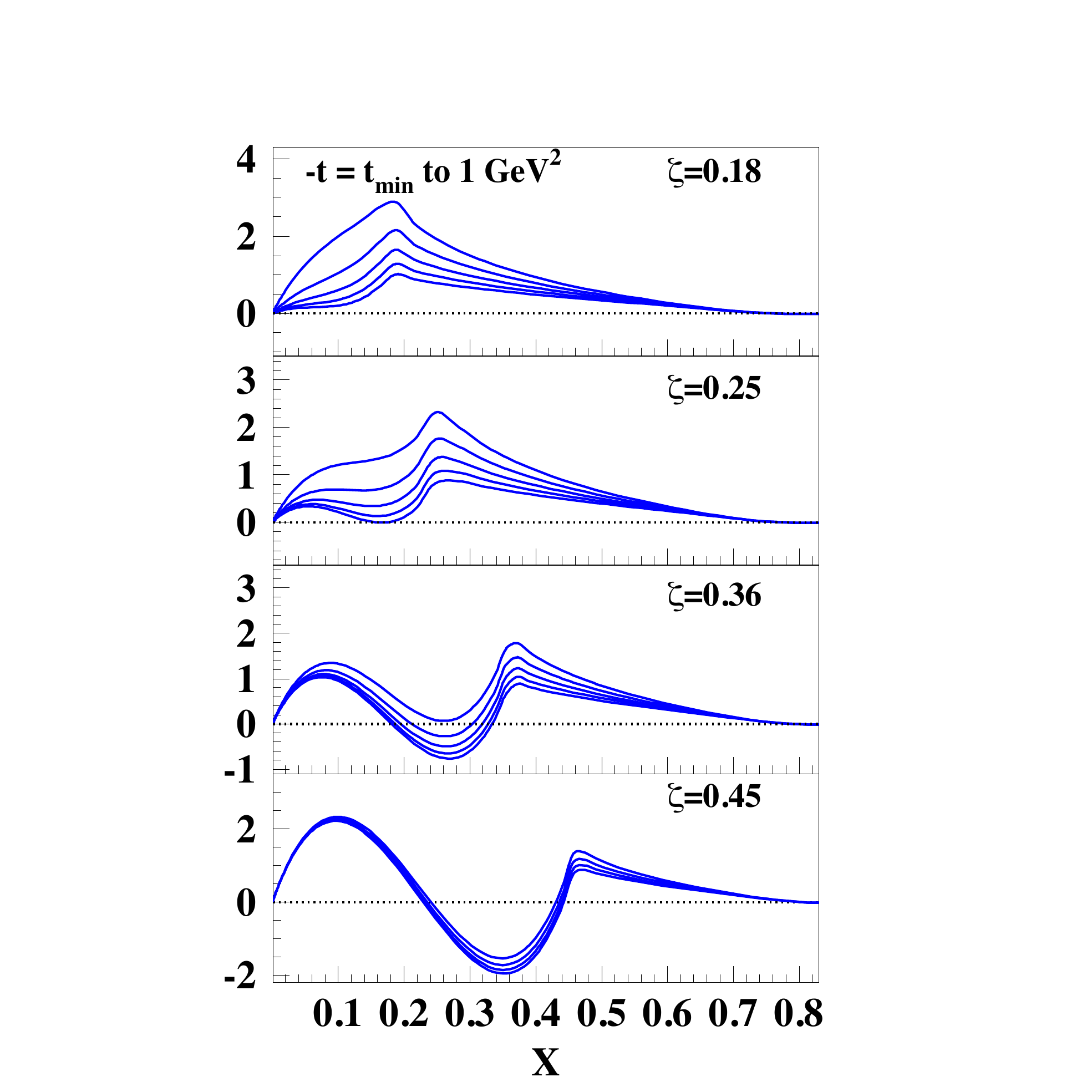}
\caption{$H_u(X,\zeta,t;Q^2)$ evaluated at $Q^2 = 2$ GeV$^2$. Each panel shows $H_u$ plotted vs. $X$ at different values of $\zeta=0.18, 0.25, 0.36, 0.45$. For each value of $\zeta$ several curves are shown that correspond to a range of values in $-t$ from $t_{min}=-M^2\zeta^2/(1-\zeta)$ to  $1$ GeV$^2$.
\label{fig7} }
\end{figure}
\begin{figure}
\includegraphics[width=9.cm]{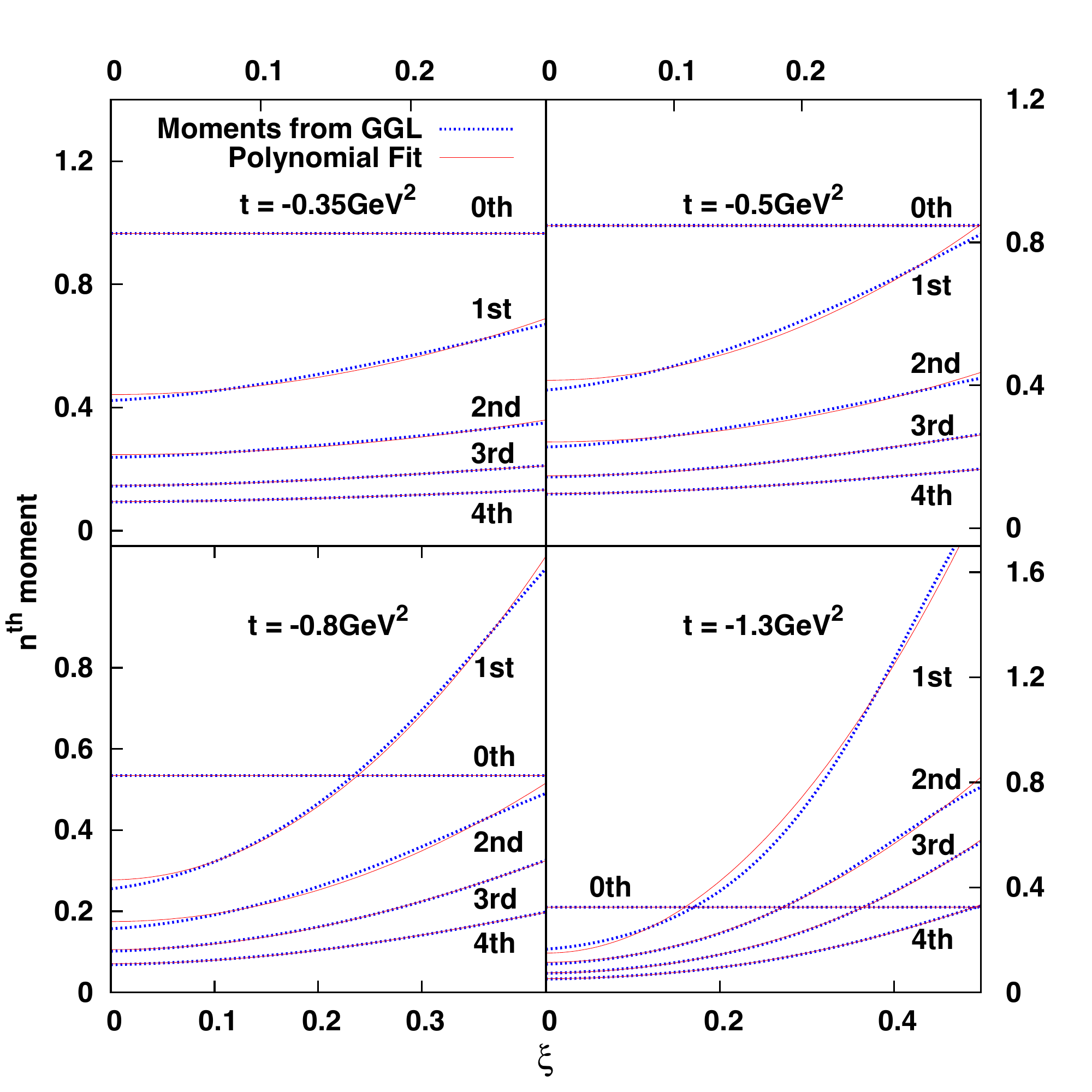}
\caption{The polynomiality property in our parametrization.  
\label{fig7_extra} }
\end{figure}

The GPDs $H_{u,d}(X,0,0)$, $E_{u,d}(X,0,0)$, $\widetilde{H}_{u,d}(X,0,0)$ are shown in Fig.\ref{fig6} both at the initial scale, $Q_o^2$, and evolved to $Q^2=$ 2 GeV$^2$. A comparison with results on $H^q$ and $E^q$ from \cite{AHLT1,AHLT2} at the scale $Q_o^2$, is also shown.  
In Fig.\ref{fig7} we show  $H_{u,d}(X,\zeta,t;Q^2)$ evaluated at $Q^2= 2$ GeV$^2$ and for a variety of ranges in $\zeta \equiv x_{Bj}$ and $Q^2$. 
In Fig.\ref{fig7_extra} we show the working of the property of polynomiality. This is, in a nutshell, a direct consequence of extending the Operator Product Expansion (OPE) 
to the off-forward case \cite{Ji1,JiLeb}, according to which the Mellin moments of GPDs read (see also \cite{Hag07})
\begin{eqnarray}
&&H_n^q(\xi,t)  =  \int^{1}_{-1} H_q(x,\xi,t) \, x^{n-1} dx  =    \nonumber \\
&=&  \! A_{n0}(t) + A_{n,2}(t)  (2 \xi)^2 + ...+ A_{n,n-1}(t) (2 \xi)^{n-1}  
\end{eqnarray}
where $n\geq1$, and only even powers of $\xi$ are included. Similar results hold for $E_q, \widetilde{H}_q, \widetilde{E}_q$ \cite{Hag07};
for $n >1$ the equation is also $Q^2$ dependent. To illustrate polynomiality, the moments of $H_u$  were plotted vs. $\xi$ at the initial scale, $Q^2_o$, for different values of $t$, and for  $n \leq 5$ (the trend shown in the figure holds for even larger values of $n$). The two sets of curves represent the calculation using the parameterization from this paper, and the results of a polynomial fit in $\xi^2$.  Clearly, our parameterization satisfies the property of polynomiality although this cannot be inferred directly from the functional form in Eq.(\ref{fit_form}). 

To summarize, we proceeded through two steps of our recursive fit. We used the flexible model described in Section \ref{sec2} to first fit the nucleon unpolarized and polarized PDFs, respectively, by using all parameters that enter the expressions at $t=0$ and $\zeta=0$. We subsequently  fitted the nucleon electromagnetic and electroweak form factors using the additional parameters that enter at $t \neq 0$ and $\zeta=0$.
At this stage of the analysis we established that in order to obtain GPDs that are constrained by a set of parameters which is consistent with the ones used for independent fits of the DIS structure functions, and of the nucleon form factors, 
a total number of $4$ (DIS) plus  $3$ (Elastic) parameters per quark flavor, per GPD is needed. 
The number of parameters is, in practice, reduced because of the physically motivated ansatze introduced in our approach, as one can see by inspecting the values in Table \ref{table_DIS}. Physical assumptions are both useful to understand the trend of data and at the same time they are known to introduce a bias. This aspect is well known to affect all hard processes multi-variable analyses, and it can be perhaps overcome in suitable neural network based approaches \cite{Forte,SOMPDF}.  

The third set of experimental constraints is given by DVCS-type data. In our analysis we use strictly DVCS data that are cleaner from the theoretical point of view, as compared to Deeply Virtual Meson Production. By fitting the CFFs that are functions of $\zeta$, $t$ and $Q^2$, we will be able to provide a constraint on the otherwise elusive 
$\zeta$ dependence of the GPDs.  
To understand how many extra parameters are needed in extending the fit to its third phase, we devised a procedure explained in the following Section.

\section{Implementation of DVCS Data}
\label{sec4}
We now discuss our procedure to extract GPDs from available DVCS data. 
Fully quantitative studies were performed in \cite{GuiMou,Mou}
where a number of observables were fitted, that were obtained from experimental data for the process $e \, p \rightarrow e p \gamma$ from both Hermes and Jefferson Lab.  
In order to have a consistent comparison, we included in our analysis a similar set of observables.
Below we list their expressions in terms of the CFFs  displayed in Sections \ref{sec2} and \ref{sec3}. In order to proceed
we first introduce the helicity formalism. This method allows us to obtain a physical interpretation of the various structures, and azymuthal angular dependences  
involved in terms of the photons helicity states.    

\subsection{Observables}
The observables included in our fit are from all the DVCS measurements that were available to us, to date. These are: the cross section for unpolarized electron scattering,  
$d \sigma/d \Phi$ \cite{HallA,HallA_n}, 
the Beam Spin Asymmetry (BSA), $A_{LU}$ \cite{HallA,HallB},
the Beam Charge Asymmetry (BCA), $A_C$ \cite{HERMES1}, 
and the Transverse Spin Asymmetries (TSAs), $A_{UT}^{DVCS}$, and $A_{UT}^{I}$ \cite{HERMES1,HERMES2}. 
The cross section for scattering of an electron/positron beam with polarization $h$ off a proton target is evaluated considering 
the sum of the DVCS and Bethe-Heitler (BH) amplitudes,
\begin{equation}
\frac{d^5 \sigma^{h}}{d Q^2 d x_{Bj}  dt  d \phi d \phi_N} = \Gamma \mid T_{BH} + T_{DVCS} \mid^2,
\end{equation}
where the factor $\Gamma$ is given by
\[ \Gamma = \frac{\alpha^3}{16 \pi^2} \frac{x_{Bj} y}{Q^2(2Mx_{Bj} \epsilon_1)} \frac{1}{\sqrt{1 + \epsilon^2}} \]
with $y=\nu/\epsilon_1$, $\epsilon_1$ being the initial electron energy, and $\nu=\epsilon_1-\epsilon_2$ the momentum transfer; $\epsilon=4M^2 x_{Bj}^2/Q^2$; 
$\phi$ is the (azymuthal) angle between the hadronic and leptonic planes in the frame where the virtual photon's momentum is  along the $z$-axis \cite{BKM}.
The unobserved helicities have been summed over, implicitly.
\footnote{Comparisons with experiments require a rotation of the angles given above from the BKM frame to either the Lab frame or the $ep$ CMF corresponding respectively to Jefferson Lab and Hermes experiments.}  
The various observables that we consider are written as
\begin{widetext}
\begin{eqnarray}
\label{dsigma}
\frac{d\sigma}{d \Phi} & = & \frac{1}{2} \left[\frac{d^4 \sigma^\uparrow}{d \Phi}  + \frac{d^4 \sigma^\downarrow}{d \Phi}  \right]=  \Gamma \left[ \mid T_{BH} \mid^2  +  \mid T_{DVCS} \mid^2 + \, \left( I^\uparrow + I^\downarrow \right) \right], \\
\label{ALU}
A_{LU} &  = &  \frac{\displaystyle \frac{d^4 \sigma^\uparrow }{d \Phi} - \displaystyle\frac{d^4 \sigma^\downarrow}{d \Phi} }{\displaystyle\frac{d^4 \sigma^\uparrow}{d \Phi} + \frac{d^4 \sigma^\downarrow}{d \Phi} }  
= \frac{\Gamma (I^\uparrow - I^\downarrow)}{2 \, d\sigma/d \Phi}, \\
\label{AC}
A_C & = &  \frac{\displaystyle \frac{d^4 \sigma^+ }{d \Phi} - \displaystyle\frac{d^4 \sigma^-}{d \Phi} }{\displaystyle\frac{d^4 \sigma^+}{d \Phi} + \frac{d^4 \sigma^-}{d \Phi} } 
= \frac{\Gamma (I^\uparrow + I^\downarrow)}{2 \, d\sigma/d\Phi}, \\
\label{AUT}
A_{UT}^{DVCS} & = & \frac{1}{S_\perp} \frac{\left( \displaystyle \frac{d^4 \sigma^+_{\Leftarrow}}{d \Phi} - \displaystyle \frac{d^4 \sigma^+_{\Rightarrow}}{d \Phi} \right) + \left(\displaystyle\frac{d^4 \sigma^{-}_{\Leftarrow}}{d \Phi} - \displaystyle\frac{d^4 \sigma^{-}_{\Rightarrow}}{d \Phi} \right)}{\displaystyle\frac{d^4 \sigma^+}{d \Phi} + \frac{d^4 \sigma^-}{d \Phi} } 
=  \frac{\Gamma \mid T^{DVCS}_{TP} \mid^2}{2 \, d\sigma/d\Phi},
 \\
A_{UT}^{I} & = & \frac{1}{S_\perp} \frac{\left( \displaystyle \frac{d^4 \sigma^+_{\Leftarrow}}{d \Phi} - \displaystyle \frac{d^4 \sigma^+_{\Rightarrow}}{d \Phi} \right) - \left(\displaystyle\frac{d^4 \sigma^{-}_{\Leftarrow}}{d \Phi} - \displaystyle\frac{d^4 \sigma^{-}_{\Rightarrow}}{d \Phi} \right)}{\displaystyle\frac{d^4 \sigma^+}{d \Phi} + \frac{d^4 \sigma^-}{d \Phi} } =
 \frac{\Gamma I_{TP} }{2 \, d\sigma/d \Phi}
\end{eqnarray}  
\end{widetext}
where $d\Phi = d \phi d x_{Bj} dt d Q^2$; the superscripts $+(-)$ refer to the beams' charge, $\uparrow(\downarrow)$ are for oppositely polarized electron beams, the subscripts $\Leftarrow(\Rightarrow)$ represent the transverse target polarizations, corresponding to the angles $\phi_S$ and $\phi_S +\pi$, respectively. 
The subscript $TP$ follows the notation of \cite{BKM} for transverse polarized target; 
the subscript $LU$ is for a longitudinally polarized beam, $L$, and an unpolarized target, $U$, while $UT$ is for an unpolarized beam, and a transversely polarized target. 

An expression for $T_{BH}$, the amplitude for the Bethe-Heitler (BH) process is given in Ref.\cite{BKM}.  
%
Here we write the amplitude in helicity basis in order to facilitate the expansion of the observables in bilinear products of amplitudes and GPDs. The basic form of the BH amplitude is
\begin{eqnarray}
T_{BH}^{h,\Lambda,\Lambda^\prime,\Lambda^\prime_\gamma} = L_{\mu\kappa}^h \epsilon^{*\, \kappa \, \Lambda_\gamma^\prime}\frac{1}{Q^2} J^{\mu}_{\Lambda,\Lambda^\prime},
\label{BH1}
\end{eqnarray}
where
\begin{eqnarray}
 L_{\mu\kappa}^h &=&  \bar{u}(k_2,h)\left[ \gamma_\mu (\gamma_\rho [k_1^\rho - q^\rho])^{-1} \gamma_\kappa \right. \nonumber \\   
 & & +\left. \gamma_\kappa (\gamma_\rho [k_2^\rho + q^\rho])^{-1} \gamma_\mu\right] u(k_1,h) 
 \label{Lmu}
 \end{eqnarray}
 and 
 \begin{eqnarray}
 J^{\mu}_{\Lambda,\Lambda^\prime} &=& \bar{U}(p^\prime,\Lambda^\prime)\left[F_1(\Delta^2) \gamma^\mu \right. \nonumber \\ 
  & & + \left. i\frac{\sigma^{\mu,\tau} \Delta_\tau}{2M}F_2(\Delta^2)\right] U(p,\Lambda).
\label{BH2}
\end{eqnarray}
The DVCS amplitude for scattering of a lepton with spin $h$ is given by 
\begin{eqnarray}
T_{DVCS}^{\Lambda,\Lambda^\prime,\Lambda^\prime_\gamma} = \bar{u}(k_2,h) \gamma_\mu u(k_1,h)  \frac{1}{Q^2}  T^{\mu \nu}_{\Lambda,\Lambda^\prime}  
\epsilon^{* \, \Lambda^\prime_\gamma}_\nu,
\label{TDVCS_1}
\end{eqnarray}
where $u(k_{1(2)},h)$ are the initial and final lepton spinors; the hadronic tensor, $T^{\mu \nu}_{\Lambda,\Lambda^\prime} $, was defined in Section  \ref{sec2}, and $\epsilon^{*\, \Lambda_\gamma}_\nu$ is
the outgoing photon polarization vector.
$T_{DVCS}$ can be expressed in terms of helicity amplitudes by considering the following expansion on the polarization vectors basis \cite{Kro96,Guichon}, 
\begin{eqnarray}
T_{DVCS}^{\Lambda,\Lambda^\prime,\Lambda^\prime_\gamma}&=&   A^+ \, f_{+,\Lambda;\Lambda_\gamma^\prime,\Lambda^\prime}
+ A^-  \, f_{-,\Lambda;\Lambda_\gamma^\prime,\Lambda^\prime}
\nonumber \\
& & +  \frac{\sqrt{Q^2}}{\nu}  A_3   \, f_{0,\Lambda;\Lambda_\gamma^\prime,\Lambda^\prime}
\label{TDVCS_3}
\end{eqnarray}
with
\begin{subequations}
\begin{eqnarray}
 A_{\pm} & = & \frac{\pm1}{\sqrt{Q^2}} \left( \sqrt{\frac{1+\epsilon}{2(1-\epsilon)}} \mp h \sqrt{2}\right) e^{\pm i \phi} \nonumber \\
 & = & \mp \frac{1}{\sqrt{2}} \bar{u}(k_2,h) [ \gamma_1 \pm i \gamma_2 ] u(k_1,h) \\
 \frac{\sqrt{Q^2}}{\nu} A_3 & = & \frac{\nu}{Q^2} \sqrt{\frac{2 \epsilon}{1-\epsilon}} \nonumber \\
& = &   \bar{u}(k_2,h) \gamma_3 u(k_1,h) 
\end{eqnarray}
\end{subequations}
 
The $f$ amplitudes were given in terms of CFFs in Section II. At LO, by disregarding the longitudinal photon polarization, the only amplitudes that were found
to contribute are: $f_{++,++}, f_{-+,-+}, f_{++,+-}, f_{-+,--}$. As a consequence, the only term contributing to the unpolarized term, $\mid T_{DVCS}\mid^2$, corresponds to the 
transverse cross section, $d \sigma_T/dt$, and it is given by
\begin{eqnarray}
\mid T_{DVCS} \mid^2& = &   \frac{1}{Q^2} \frac{1}{2(1-\epsilon)} (\mid f_{++,++}\mid^2 +   \mid f_{-+,-+}\mid^2+  \nonumber \\
& & \mid f_{++,+-}\mid ^2 +  \mid f_{-+,--}\mid^2 )
\end{eqnarray}
In terms of CFFs,
\begin{widetext}
\begin{eqnarray}
\label{TDVCS_GPD}
\mid T_{DVCS} \mid^2& = &  \frac{1}{Q^2} \frac{1}{2(1-\epsilon)} \frac{1}{(2-x_{Bj})^2} \left[ 
4 (1-x_{Bj}) ({\cal H}^*{\cal H} + \widetilde{{\cal H}}^* \widetilde{{\cal H}}) 
- x_{Bj}^2({\cal E}^*{\cal H}  + {\cal H}^*{\cal E}  + \widetilde{\cal E}^*\widetilde{\cal H}  +\widetilde{\cal H}^*\widetilde{\cal E} ) -  \right. \nonumber \\ 
 &  - & \left. 4 \left( \frac{x_{Bj}^2}{1-x_{Bj}} + \frac{t}{4M^2} \right) {\cal E}^*{\cal E} - 4 x_{Bj}^2 \frac{t}{4M^2} \widetilde{\cal E}^*\widetilde{\cal E}     \right] 
\end{eqnarray}
\end{widetext}
Eq.(\ref{TDVCS_GPD})  is analogous to term $c_0^{DVCS}$ in the expansion given in \cite{BKM}. Note that the sum over all the unobserved helicities is implicit.

An analogous decomposition into the virtual photon polarization basis for the Bethe-Heitler amplitude of Eq.(\ref{BH1}) has the form
\begin{eqnarray}
T_{BH}^{h,\Lambda,\Lambda^\prime,\Lambda^\prime_\gamma} & = & B_{h, \Lambda_\gamma^\prime}^+J_{+, \Lambda ; \Lambda^\prime} + B_{h, \Lambda_\gamma^\prime}^-J_{-, \Lambda ; \Lambda^\prime}  \nonumber \\
& + &\frac{\sqrt{Q^2}}{\nu} B_{h, \Lambda_\gamma^\prime}^0J_{0, \Lambda ; \Lambda^\prime}.
 \label{BH_hel}
 \end{eqnarray}
The hadronic amplitudes are
\begin{eqnarray}
J_{\pm1, \Lambda ; \Lambda^\prime=\Lambda}&=& -\frac{p_\perp^\prime}{\sqrt{1-\xi^2}}F_2(t) \nonumber \\
J_{\pm1, \Lambda ; \Lambda^\prime= - \Lambda}&=&\Lambda \frac{(p_\perp^\prime)^2}{2M\sqrt{1-\xi^2}}F_2(t) \nonumber \\
 & + & \delta_{\Lambda,\pm} \frac{\sqrt{2}M\xi}{\sqrt{1-\xi^2}}(F_1(t)+F_2(t)),
 \label{Jmu}
 \end{eqnarray}
 where terms of order $(\Delta_\perp/p^+)^2$ and $(M/p^+)^2$ were dropped.

 The lepton tensor for the Bethe-Heitler amplitude can be calculated from Eq.(~\ref{Lmu}) using the relation
\begin{equation}
\gamma^\mu \gamma^\nu \gamma^\rho = g^{\mu\nu} \gamma^\rho +g^{\nu\rho} \gamma^\mu -g^{\rho\mu} \gamma^\nu -i\epsilon^{\mu\nu\rho\sigma}\gamma_\sigma \gamma_5.
\label{3gamma}
\end{equation}
Momentum conservation gives the exchanged virtual photon momentum as $q_X=k_2 + q^\prime -k_1 = q^\prime - q = q^\prime -(p-p^\prime) = q^\prime - \Delta$. It can be seen that the Dirac algebra is reduced to evaluating a single $\gamma^\sigma$ or $\gamma^\sigma\gamma^5$. 
 
The interference term is given by,
\begin{widetext}
\begin{eqnarray}
\label{interf}
I & = & \sum_{h,\Lambda \Lambda^\prime \Lambda^\prime_\gamma} \left(T_{BH}^{* \, h,\Lambda \Lambda^\prime \Lambda^\prime_\gamma} T_{DVCS}^{h,\Lambda \Lambda^\prime \Lambda^\prime_\gamma}
+ T_{DVCS}^{* \, h,\Lambda,\Lambda^\prime,\Lambda^\prime_\gamma} T_{BH} ^{h,\Lambda \Lambda^\prime \Lambda^\prime_\gamma} \right).
\end{eqnarray}
We consider the following expansion over helicity states of $T_{BH}^{h,\Lambda \Lambda^\prime \Lambda_\gamma}$ and  $T_{DVCS}^{h,\Lambda,\Lambda^\prime,\Lambda^\prime_\gamma}$ in terms of $f_{\Lambda_\gamma, \Lambda; \Lambda_\gamma^\prime, \Lambda^\prime}$ for fixed lepton helicity $h$,
\begin{eqnarray}
I^h  & = &  
\left[ \left(\left[B_{h,+1}^{+1*}J_{+1; +,+} + B_{h,+1}^{0*}J_{0; +,+}+B_{h,+1}^{-1*}J_{-1; +,+}\right] A_h^{+1} \right. \right. \nonumber \\
 & & \left. + \left[ B_{h,-1}^{+1*}J_{-1; +,+} + B_{h,-1}^{0*}J_{0; +,+} + B_{h,-1}^{-1*}J_{+1; +,+} \right] A_h^{-1} \right) f_{+,+;+,+}  \nonumber  \cr
 & & + \left( \left[ B_{h,-1}^{+1*}J_{+1; +,+} + B_{h,-1}^{0*}J_{0; +,+}+B_{h,-1}^{-1*}J_{-1; +,+}\right] A_h^{-1} \right. \cr
 & & + \left. \left[ B_{h,+1}^{+1*}J_{-1; +,+} + B_{h,+1}^{0*}J_{0; +,+}+B_{h,+1}^{-1*}J_{+1; +,+}\right] A_h^{+1} \right) f_{-,+;-,+} \nonumber \cr
  &  & +  
\left[ \left(\left[B_{h,+1}^{+1*}J_{+1; +,-} + B_{h,+1}^{0*}J_{0; +,-}+B_{h,+1}^{-1*}J_{-1; +,-}\right] A_h^{+1} \right. \right. \nonumber \\
 & & \left. +\left[B_{h,-1}^{+1*}J_{+1; -,+} - B_{h,-1}^{0*}J_{0; +,-}-B_{h,-1}^{-1*}J_{+1; +,-}\right] A_h^{-1} \right) f_{+,+;+,-}   \cr
 & & + \left( \left[ B_{h,-1}^{+1*}J_{+1; +,-} + B_{h,-1}^{0*}J_{0; +,-}+B_{h,-1}^{-1*}J_{-1; +,-}\right] A_h^{-1} \right. \cr
 & & +\left. \left. \left[ B_{h,+1}^{+1*}J_{+1; -,+} - B_{h,+1}^{0*}J_{0; +,-}-B_{h,+1}^{-1*}J_{+1; +,-}\right] A_h^{+1} \right) f_{-,+;-,-} \right]
\end{eqnarray}
where the various helicity dependent terms are defined in Eq.(\ref{Lmu},\ref{Jmu},\ref{TDVCS_3}).

Eq.(\ref{interf}) then has the following structure analogous to the leading terms in the expansion in \cite{BKM},
\begin{equation}
I \propto c_0^I + c_1^I \cos \phi + h s_1^I \sin \phi 
\end{equation}
with coefficients given by
\begin{eqnarray}
c_0^{I} & = & - 8 \frac{(2 - 2y)^3}{1-y}  K^2    \left[ F_1  \Re e{\cal H}  + \frac{x_{Bj}}{2-x_{Bj}}  (F_1 + F_2)  \Re e\widetilde{\cal H} - \frac{t}{4M^2} F_2 \Re e{\cal E} 
\right]  
\end{eqnarray}
\begin{eqnarray}
c_1^{I} & = & - 8(2 - 2y + y^2) K    \left[ F_1  \Re e{\cal H}  + \frac{x_{Bj}}{2-x_{Bj}}  (F_1 + F_2)  \Re e\widetilde{\cal H} - \frac{t}{4M^2} F_2 \Re e{\cal E} 
\right]  
\end{eqnarray}
\begin{eqnarray}
s_1^{I} & = & 8y(2-y) K  \left[ F_1  \Im m{\cal H}  + \frac{x_{Bj}}{2-x_{Bj}}  (F_1 + F_2)  \Im m\widetilde{\cal H} - \frac{t}{4M^2} F_2 \Im m{\cal E} 
\right] 
\end{eqnarray}
The kinematical factor $K$ is, at leading order in $-t/Q^2$,
\[ K = \left[ \frac{-t}{Q^2}  \left(1-x_{Bj}\right) \left(1 - y - \frac{y^2 \epsilon^2}{4} \right) \left(1 - \frac{t_{min}}{t} \right) \left( 1 + \epsilon\right)^{1/2}  \right]^{1/2} \]  
with $t_{min} = - x_{Bj}^2/(1-x_{Bj}) M^2$. 
\end{widetext}

For a practical approach we streamline both the kinematical dependence and the GPD content of the various observables. By keeping
the leading terms in $-t/Q^2$, and in $x_{Bj}$, we obtain 
\begin{widetext}
\begin{subequations}
\begin{eqnarray}
\label{dsigma1}
\frac{d\sigma_{e^-}}{d \Phi} 
& = & \Gamma \left( B(\phi)  + C(\phi)  \cos \phi  \right)  
\\
\label{ALU1}
A_{LU} &  = & \displaystyle \frac{ A(\phi)  \sin \phi }{B(\phi)  + C(\phi)  \cos \phi} 
 \\
\label{AC1}
A_C & = & \displaystyle \frac{ D(\phi)  \cos \phi }{B(\phi)  + C(\phi)  \cos \phi} 
\\
A_{UT}^{DVCS} & = &  \frac{ E(\phi) \sin(\phi - \phi_S)}{B(\phi)  + C(\phi)  \cos \phi} 
 \\
\label{AUTI}
A_{UT}^{I} & = & \frac{ (F(\phi) +  G(\phi)  \cos \phi)\sin(\phi-\phi_S) + H(\phi) \cos(\phi-\phi_S)  \sin \phi}{B(\phi)  + C(\phi)  \cos \phi} 
\end{eqnarray}  
\end{subequations} 
\end{widetext}
where, 
\begin{widetext}
\begin{subequations}
\label{ABC}
\begin{eqnarray}
A & = & K_I(\phi) 
\left(F_1  \, \Im m {\cal H}  + \displaystyle\frac{x_{Bj} }{2-x_{Bj} } (F_1 + F_2) \Im m {\cal{\widetilde{H}}} \right)  \\
B & = & C_0^{BH}(\phi)
+ K_{DVCS}^0 \left(   \Re e^2 {\cal H} + \Im m^2 {\cal H} + \Re e^2 {\cal{\widetilde{H}}}  + \Im m^2 {\cal {\widetilde{H}}}  \right) 
 +  K_I ^0(\phi) \left( F_1  \Re e {\cal H}  + \frac{x_{Bj} }{2-x_{Bj} } (F_1 + F_2) \Re e {\cal{\widetilde{H}}} \right)  
 \\
C & = & \left[ C_1^{BH}(\phi) + K_I^1(\phi)  \left( F_1  \Re e {\cal H}  + \frac{x_{Bj}}{2-x_{Bj}} (F_1 + F_2) \Re e{\cal {\widetilde{H}}} \right)  \right]  
\\
D & = & K_I(\phi)  \left(F_1  \, \Re e {\cal H}  + \displaystyle\frac{x_{Bj} }{2-x_{Bj} } (F_1 + F_2) \Re e {\cal {\widetilde{H}}} \right) 
\\
E & = &  K_{DVCS}^{UT, \, 0}  \left( \Re e {\cal E} \, \Im m  {\cal H}  -  \Re e {\cal H} \, \Im m {\cal E}  \right)
 \\
F & = &   K_I^{UT, \, 0}(\phi)  \left( \frac{t}{4 M^2} (2-x_{Bj}) F_1 \, 
\Im m {\cal E}  - \frac{t}{M^2} \frac{1-x_{Bj}}{2-x_{Bj}} F_2  \,  \Im m {\cal H}  \right)
\\
G & = & K_I^{UT, \, 1}(\phi)  \left( \frac{t}{4 M^2} (2-x_{Bj}) F_1 \, 
\Im m {\cal E}  - \frac{t}{M^2}\frac{1-x_{Bj}}{2-x_{Bj}} F_2  \, \Im m {\cal H}  \right)
 \\
H & = & K_I^{UT, \, 1}(\phi)    \left( -\frac{t}{4M^2} x_{Bj} F_1 \, \Im m {\cal \widetilde{E}} 
+ \frac{t}{M^2}\frac{1-x_{Bj}}{2-x_{Bj}} F_2  \, \Im m {\cal H}  \right).
\end{eqnarray}  
\end{subequations} 
\end{widetext}
$C_{0(1)}^{BH}(\phi)$ enter the BH cross section \cite{BKM}. The factors $K$ are kinematical coefficients which depend on $t, x_{Bj}, y, Q^2, \epsilon$;  for the 
BH and interference terms, they depend also on $\phi$ due to the BH propagators. 

\subsection{Fit Results}
\begin{figure}
\includegraphics[width=9.cm]{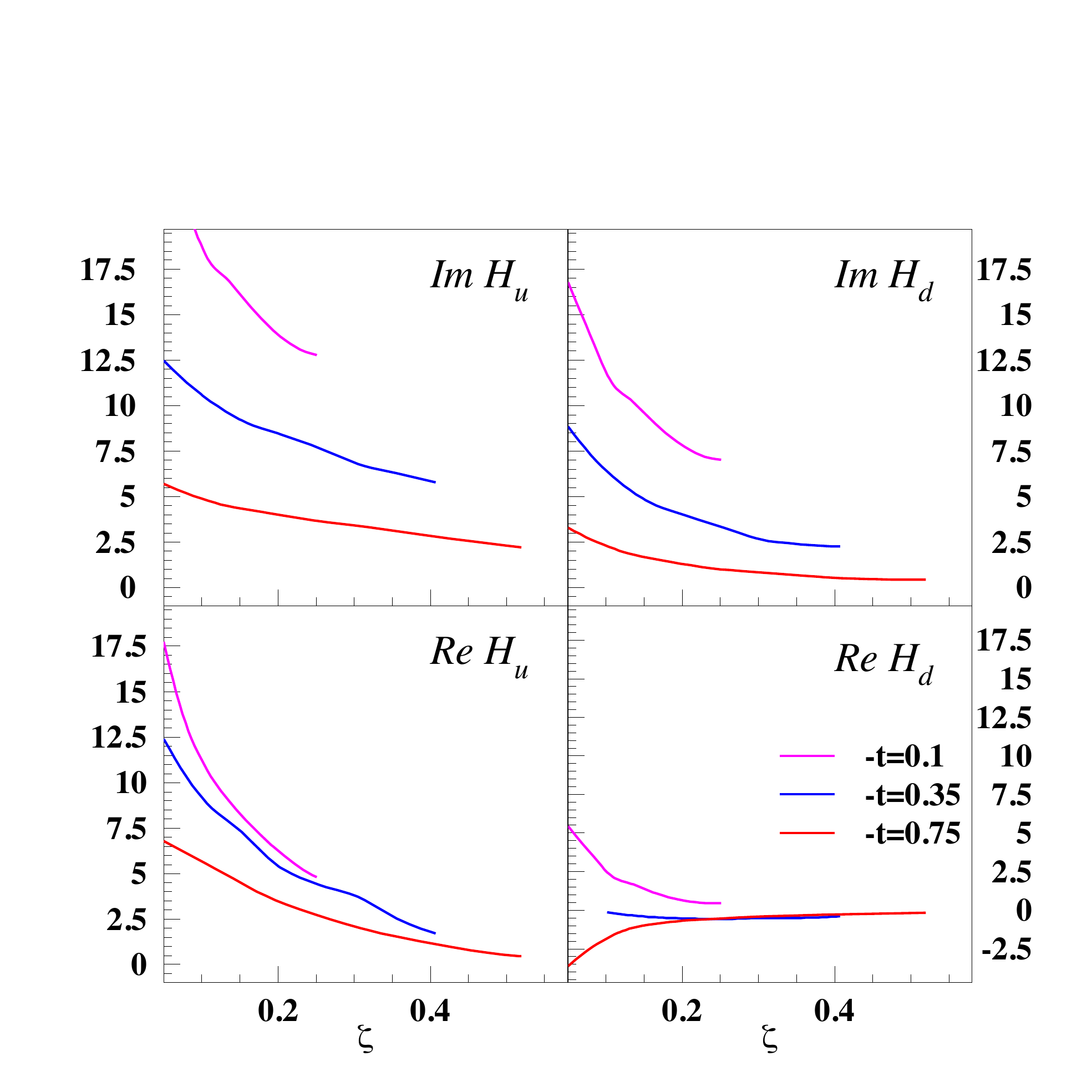}
\caption{(color online) Real and imaginary parts of the CFFs, ${\cal H}(\zeta,t)$,  Eqs.(\ref{ABC}). The CFFs are plotted vs. $x_{Bj} \equiv \zeta$, for different values of $t$, at $Q^2 = 2$ GeV$^2$. Similar results are obtained for $E$ and $\widetilde{H}$. }
\label{fig9}
\end{figure}

We present results from our fit including all parameters that were fixed using the reggeized diquark parametrization described in Sections \ref{sec2}, \ref{sec3}, and displayed in Table \ref{table_DIS}.  We introduce additional flexibility through extra parameters entering Eq.(\ref{regge}) in order to constrain the $\zeta$ dependence from all available DVCS data. Although in principle as many $\beta$ parameters as the number of flavors and GPDs could be introduced, given the small data set presently available, we use one value of $\beta$ for all GPDs. More accurate studies including flavor and GPD dependent $\beta$ parameters will be conducted as  more data become available. In Fig.\ref{fig9} we show the real and imaginary parts of the CFFs, ${\cal H}(\zeta,t)$, appearing in Eqs.(\ref{ABC}). Similar results are obtained for $E$, $\widetilde{H}$, and $\widetilde{E}$. One can see that the slope in $\zeta$ flattens out as $-t$ increases.  

Our fit uses the two currently available sets of data, from both Hall A and Hall B collaborations at Jefferson Lab, and the Hermes collaboration, respectively.  
Since the kinematical ranges covered by the two experiments only partially overlap, in this first step we start by fitting the Jefferson Lab set, and subsequently extend the
results of our fit to predict the behavior of the Hermes set.

\begin{figure}
\includegraphics[width=10.cm]{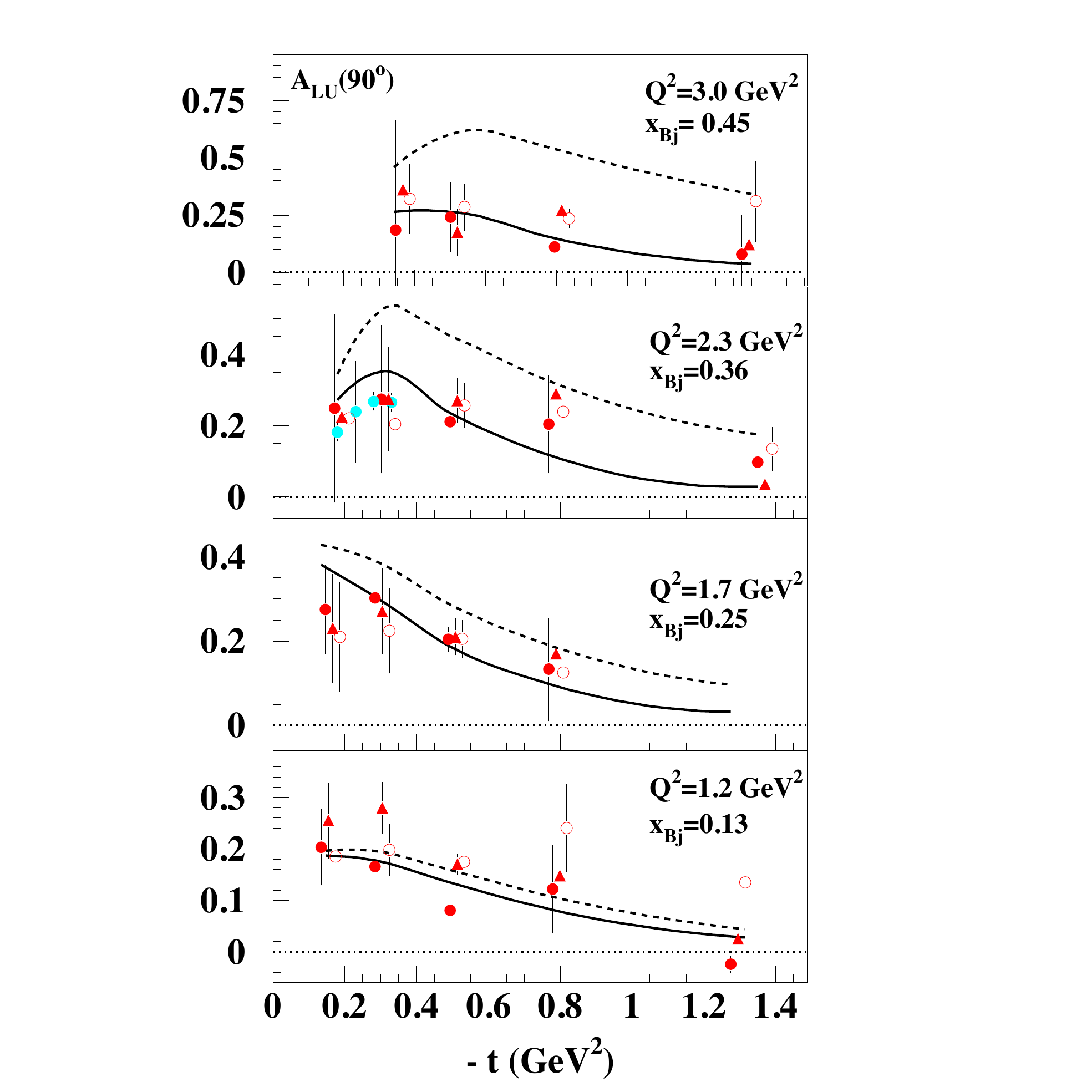}
\caption{Beam spin asymmetry, $A_{LU}(\phi=90^o)$ in 12 of the $x_{Bj}$ and $Q^2$ bins measured in Hall B \cite{HallB}. The data points were extracted
by fitting $A_{LU}(\phi)$, however they do not represent the uncertainties reported in the experimental analysis.  The second panel from the top includes also data from Hall A \cite{HallA}. The full circles, open circles, and triangles represent data in similar $x_{Bj}$ and $t$ bins, but at $Q^2$ values slightly displaced around the value reported in the legend. All curves were obtained at the kinematics displayed in the figure.  
Dashed lines: results from the fit using only the PDF and form factors constraints as from Table \ref{table_DIS}. 
The full lines represent the effect of introducing the $\zeta$ dependent term, Eq.(\ref{regge}), in the numerator of the asymmetry only. }
\label{fig10}
\end{figure}
\begin{figure}
\includegraphics[width=10.cm]{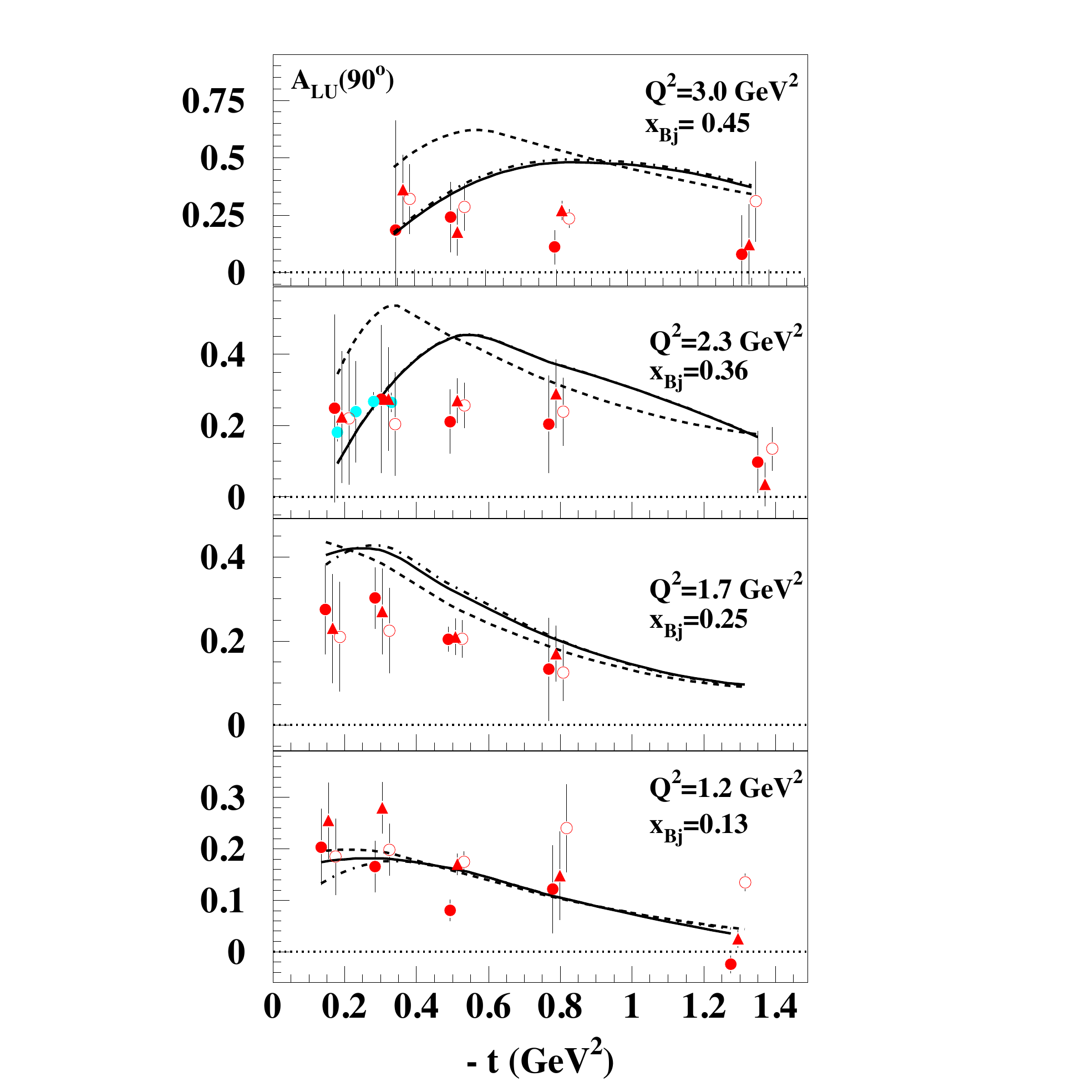}
\caption{Experimental data: same as Fig.\ref{fig10}.
Dashed lines: results from the fit using only the PDF and form factors constraints as from Table \ref{table_DIS}. 
The effect of the $\zeta$ dependent terms given in Eq.(\ref{betaI}), and Eq.(\ref{betaII}), is given by the full lines and
the dot-dashed lines, respectively.}
\label{fig10b}
\end{figure}
%

\subsubsection{Jefferson Lab Data}
The DVCS data on $A_{LU}$ from Hall B \cite{HallB}, and on the sum and the difference of the beam polarization cross sections from Hall A \cite{HallA} 
were implemented directly in our fit. This allows us to constrain the parameter $\beta$. In Figs.\ref{fig10} and \ref{fig10b} we show the results of our fits using the quantity 
$A_{LU}(90^o) = A/B$, Eq.(\ref{ALU1}). The experimental points in the figure were obtained by fitting $A_{LU}(\phi)$ in the 12 bins displayed in Ref.\cite{HallB} (figure 4). The statistical and systematic errors were added in quadrature, no error correlations were considered (this is giving rise in our case to larger error bands, although the central points coincide with the ones in Ref.\cite{HallB}).  A similar procedure was used for the Hall A data that are also displayed in the figures. 
The dashed lines in both figures are a prediction of the fit using only the PDFs and form factors constraints. Clearly by taking only these constraints, the $\zeta$ slope of the CFFs is unconstrained and evidently off the data trend, as it can be seen in the larger $\zeta \equiv x_{Bj}$ bins. The additional term in Eq.(\ref{regge}) can regulate this behavior. Two possible ways of implementing it are shown respectively by the full lines in the two figures. In Fig.\ref{fig10} a multiplicative term was considered only in the numerator  of the asymmetry, given by $A$. In Fig.\ref{fig10b} the zeta dependent factor was introduced in the GPDs, and used to calculate the CFFs in both the numerator and denominator of the asymmetry. The effect of introducing such term in the GPDs gives a different dependence that can be ascribed to modifying both the real and imaginary parts of the CFFs. We show results using two different expressions for $\beta$, Eq.(\ref{betaI}) and Eq.({\ref{betaII}), in order to illustrate some of  the subtleties that are involved in the extraction of the CFFs, and GPDs from the data. While the two expressions give almost identical results for the asymmetry, they impact the various terms, $A, B, C$ in Eqs.(\ref{ABC}) in different ways. With more data in hand, including a separation of the absolute cross sections, one will be able to perform precise fits of the behavior in $\zeta$.  

In Figs.\ref{fig11}, \ref{fig11b}, \ref{fig11c},\ref{fig11d}  we analyze the effect of the different GPD components on the fit to Hall B data. 
Fig.\ref{fig11} shows the contribution of the BH term and of the coefficients $A$, and $B$ from Eqs.(\ref{ABC}) with the $\zeta$ dependent term from Eq.(\ref{betaI}) 
(bullets),
and without it (full curves).  
\begin{figure}
\includegraphics[width=8.cm]{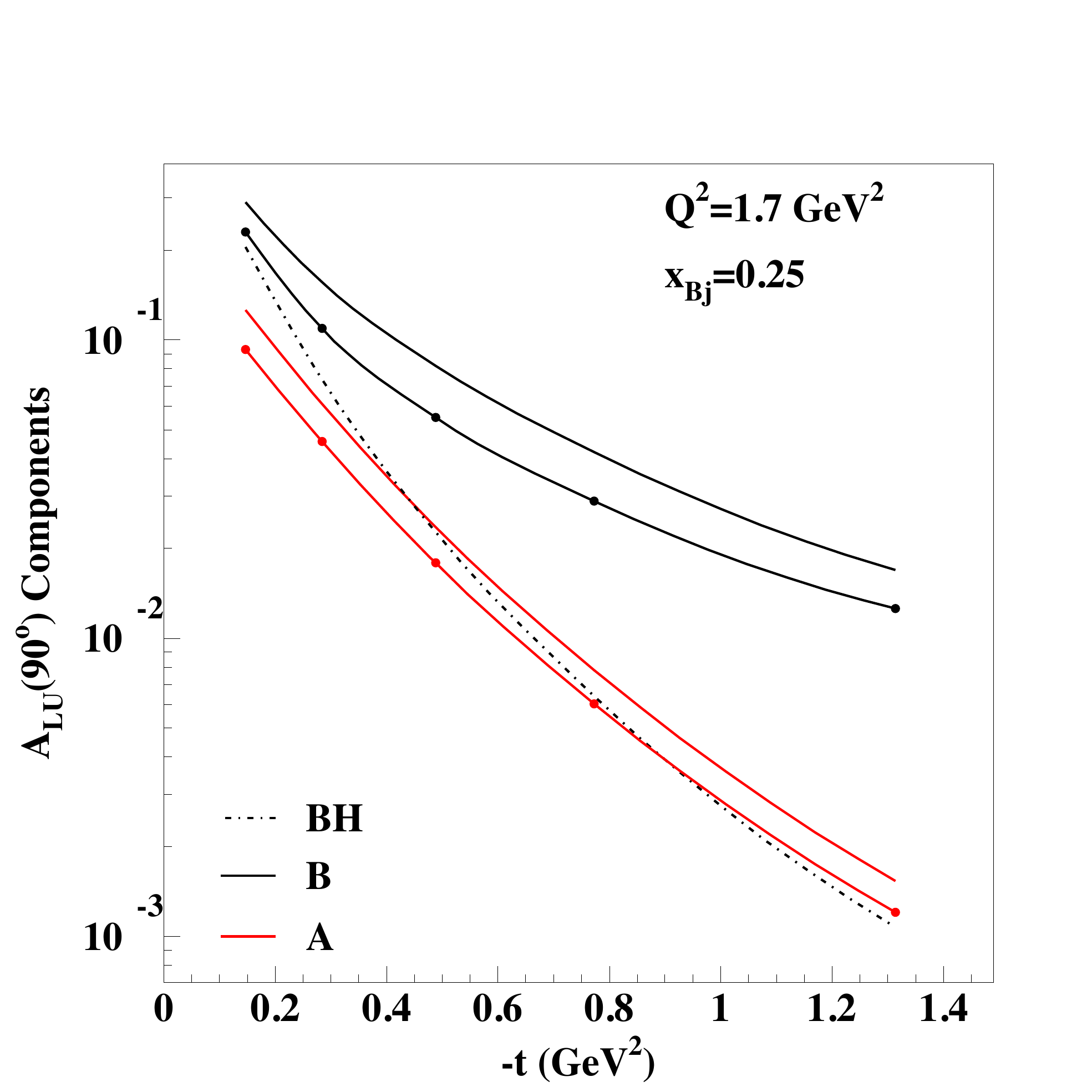}
\caption{Contribution of the BH term (dot-dashed line) and of the coefficients $A$, and $B$ from Eqs.(\ref{ABC}) with the $\zeta$ dependent term from Eq.(\ref{betaI}) 
(bullets),
and without it (full curves), at $Q^2= 1.7$ GeV$^2$ and $x_{Bj} = 0.25$. Similar results are obtained in other kinematics.}
\label{fig11}
\end{figure}
Fig.\ref{fig11b} shows the separate contributions of the numerator and denominator, $A$, and $B$, Eqs.(\ref{ABC}), to the ratio defining the asymmetry $A_{LU}(90^o)$.
$B$ is given by the sum of the three terms displayed in the figure.
\begin{figure}
\includegraphics[width=8.cm]{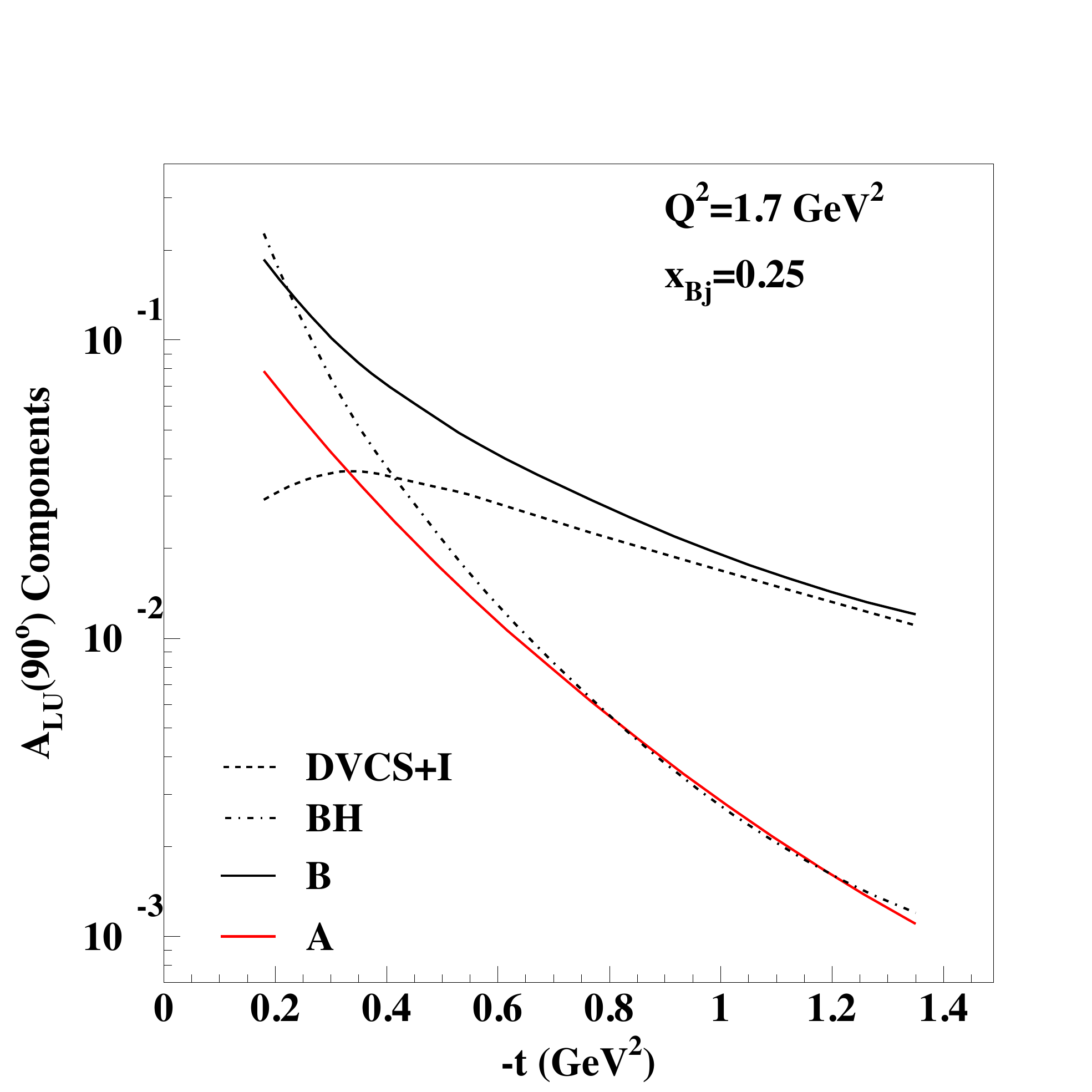}
\caption{Separate contributions of the numerator and denominator, $A$, and $B$, Eqs.(\ref{ABC}), to the $A_{LU}(90^o)$.
$B$ is given by the BH term (dot-dashes) plus the sum of the $|T_{DVCS}|^2$ and interference terms (dashes).
$Q^2= 1.7$ GeV$^2$ and $x_{Bj} = 0.25$; similar results are obtained in other kinematics.}
\label{fig11b}
\end{figure}
Notice that the real part of the CFFs enters only $T_{DVCS}^2$, which is a relatively small contribution. To understand its impact on the various observables,
in Fig.\ref{fig11c} we plotted results both including and excluding the real part.

\begin{figure}
\includegraphics[width=8.cm]{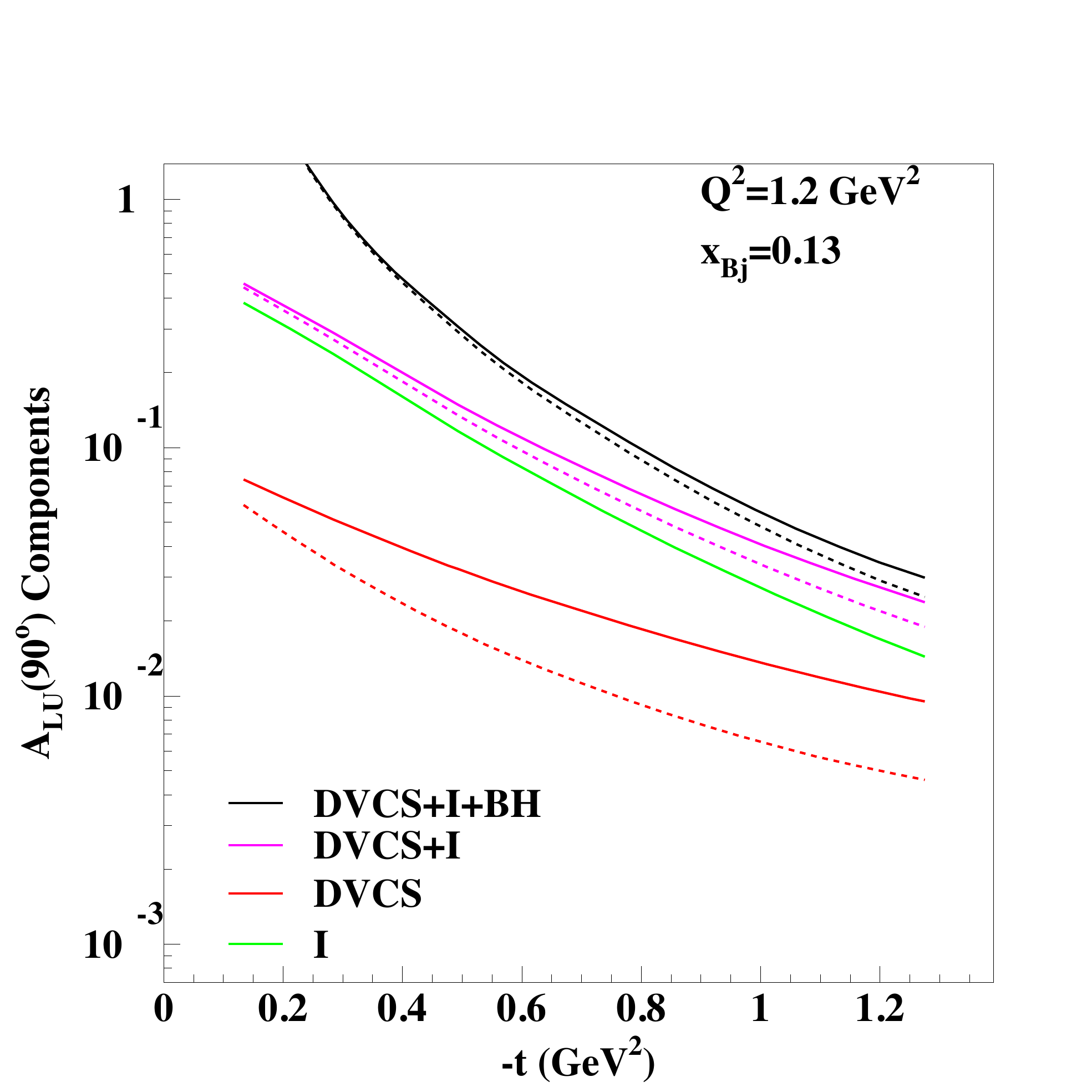}
\caption{Impact of different GPD components on data fit. The effect
of including (full curves) and excluding (dashed curves) the real part of the CFFs is shown for each component. $Q^2= 1.3$ GeV$^2$ and $x_{Bj} = 0.12$; similar results are obtained in other kinematics.}
\label{fig11c}
\end{figure}
\begin{figure}
\includegraphics[width=8.cm]{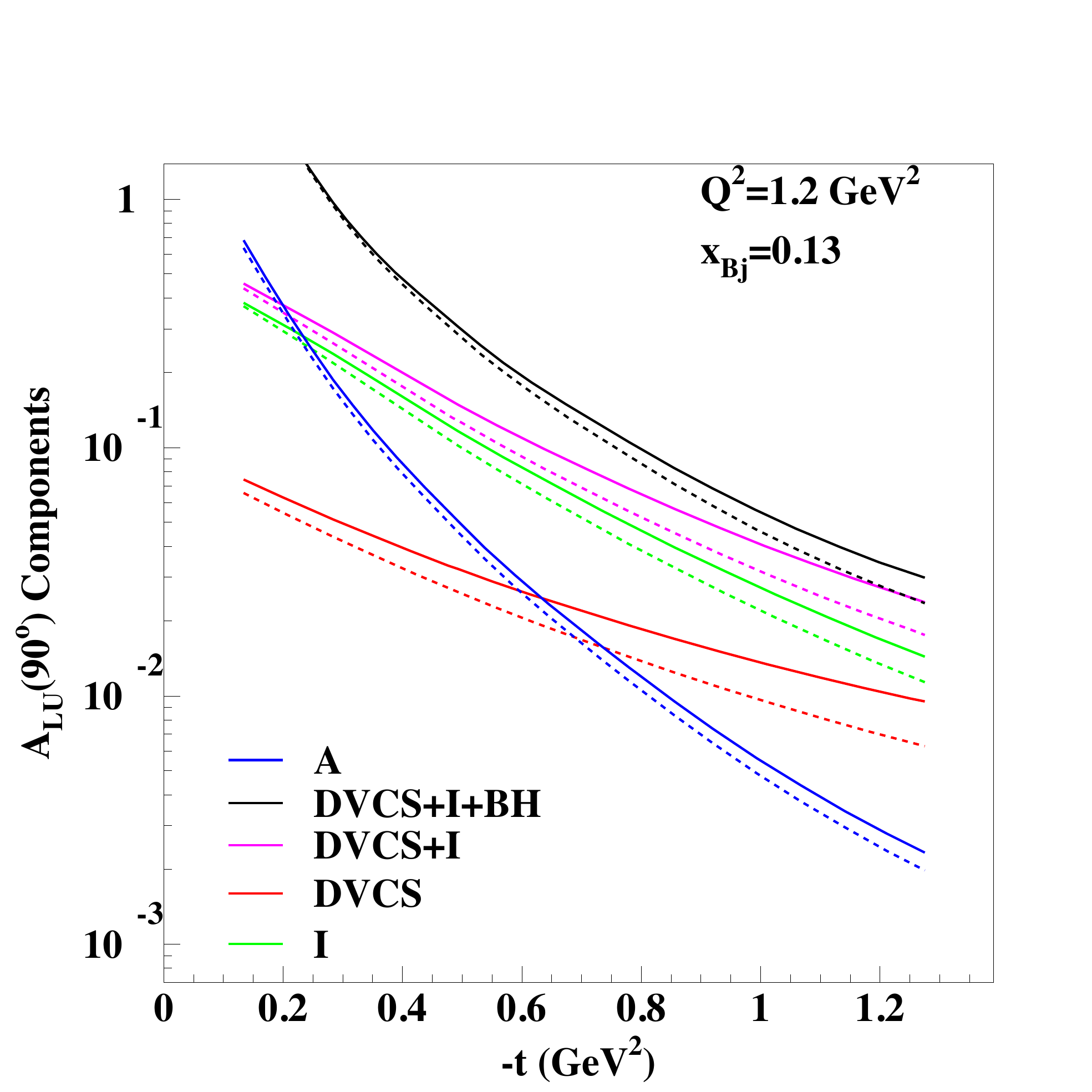}
\caption{Impact of different GPD components on data fit. The effect
of including (full curves), and excluding (dashed curves) $\widetilde{H}$ in the CFFs evaluation is shown. The contribution of $E$ is always negligible; $Q^2= 1.3$ GeV$^2$ and $x_{Bj} = 0.12$.
Similar results are found in other kinematical bins.}
\label{fig11d}
\end{figure}
Finally, in Fig.\ref{fig11d} we show the effect of the GPD $\widetilde{H}$ on the fit. We confirm the result also quantitatively reported in Ref.\cite{Mou} 
that DVCS data from an unpolarized proton target at  Jlab kinematics are dominated by the contribution of the GPD $H$. The dashed curves in the figure were obtained by disregarding the contribution of $\widetilde{H}$. $E$ and $\widetilde{E}$ have also a very little impact on the data fit.   

In Fig.\ref{fig12} we show the results of our fit vs. Hall A data \cite{HallA}. These are given as the  "sum" and  "difference" of the two polarizations for the electron beam.
Together with the data we also plot the results of a fit performed in \cite{HallA} (yellow bands). All theoretical curves are shown with and without the $\zeta$ dependent 
correcting factor from Eq.(\ref{regge}).
Moreover, while we confirm that the sum, or absolute cross section part is dominated by the BH term, we also point out the importance of the contribution from the pure DVCS scattering, or $\mid T_{DVCS}\mid ^2$, at leading order in $Q^2$. Also shown are the theoretical error bands for the asymmetry. 
%
\begin{figure}
\includegraphics[width=8.cm]{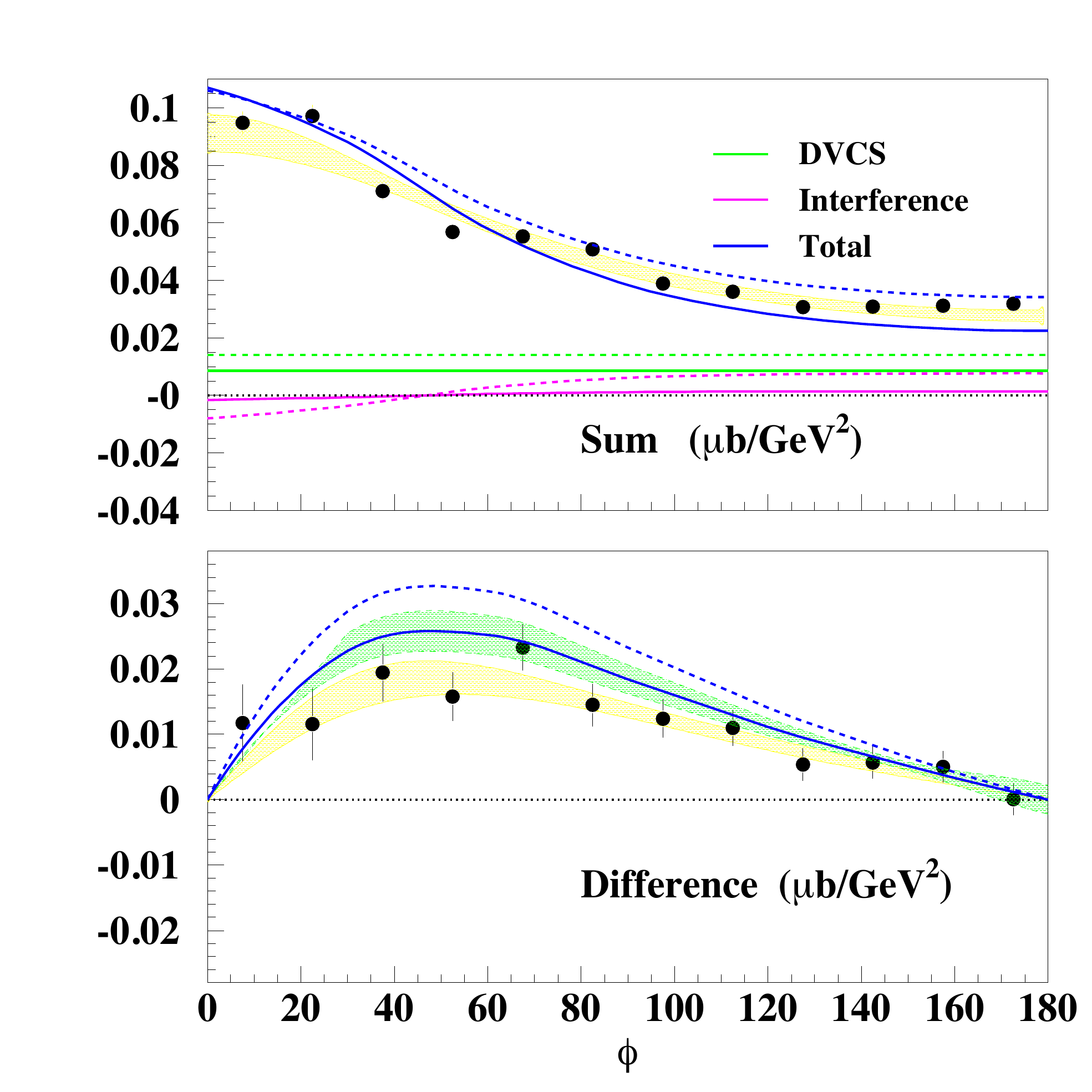}
\caption{(color online) HallA data \cite{HallA} for the "sum" (upper panel) and  "difference"  (power panel) of the two electron beam polarizations. Shown are curves including the contribution of the $\zeta$ dependent  factor from Eq.(\ref{regge}) (full lines), and neglecting it (dashed lines). All terms (DVCS, Interference and Total are shown for the "sum" graph. The yellow bands in both panels represent the error of the data fit. The green band in the asymmetry graph is the theoretical error from our parameterization.}
\label{fig12}
\end{figure}
In Fig.\ref{fig13} we  compare our results to the experimental extraction of the imaginary part of the interference term coefficient in $A_{LU}$ Eqs.(\ref{ALU1}) and (\ref{ABC}).
The role of the CFF for $\widetilde{H}$ proves fundamental in determining the slope vs. $-t$ of this term. In order to illustrate this, we show a comparison with a previous calculation of the interference term where $\widetilde{H}$ was not included \cite{AHLT2}. 
\begin{figure}
\includegraphics[width=8.cm]{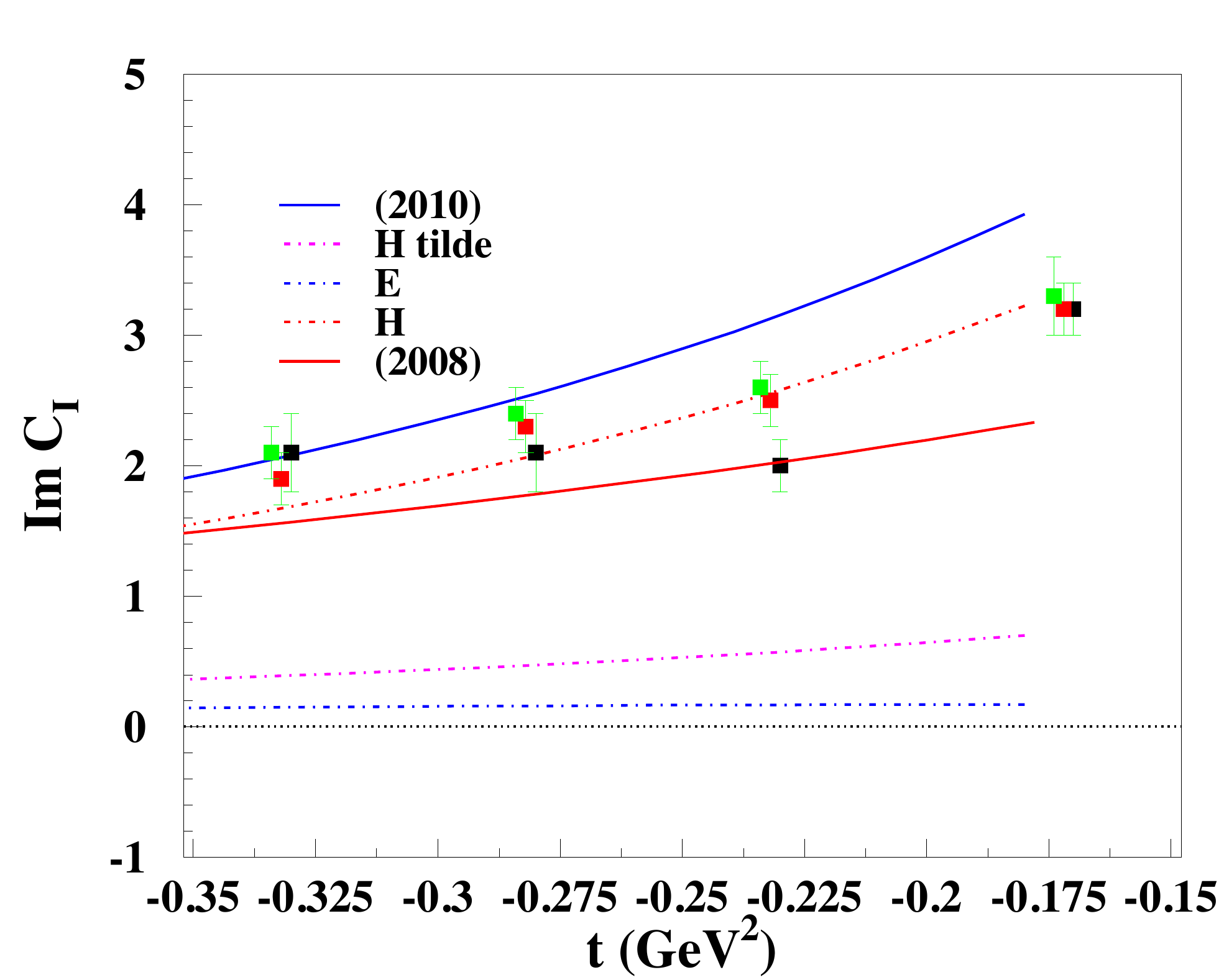}
\caption{Coefficient $C$, Eqs.(72) extracted from Hall A data \cite{HallA}. Shown are the contributions from the GPDs, $H$, $E$ and $\widetilde{H}$.
All curves include the term in Eq.(\ref{regge}). A comparison with a previous prediction based on a simplified diquark model, and including $H$ only \cite{AHLT2} 
is also shown.}
\label{fig13}
\end{figure}
\subsubsection{Hermes data}
In the second phase of our analysis, we use our fit results to Jlab data to predict the quantities, $A_{LU}$, $A_C$, and $A_{UT}$ extracted at Hermes \cite{HERMES1,HERMES2}.
\begin{figure}
\includegraphics[width=9.5cm]{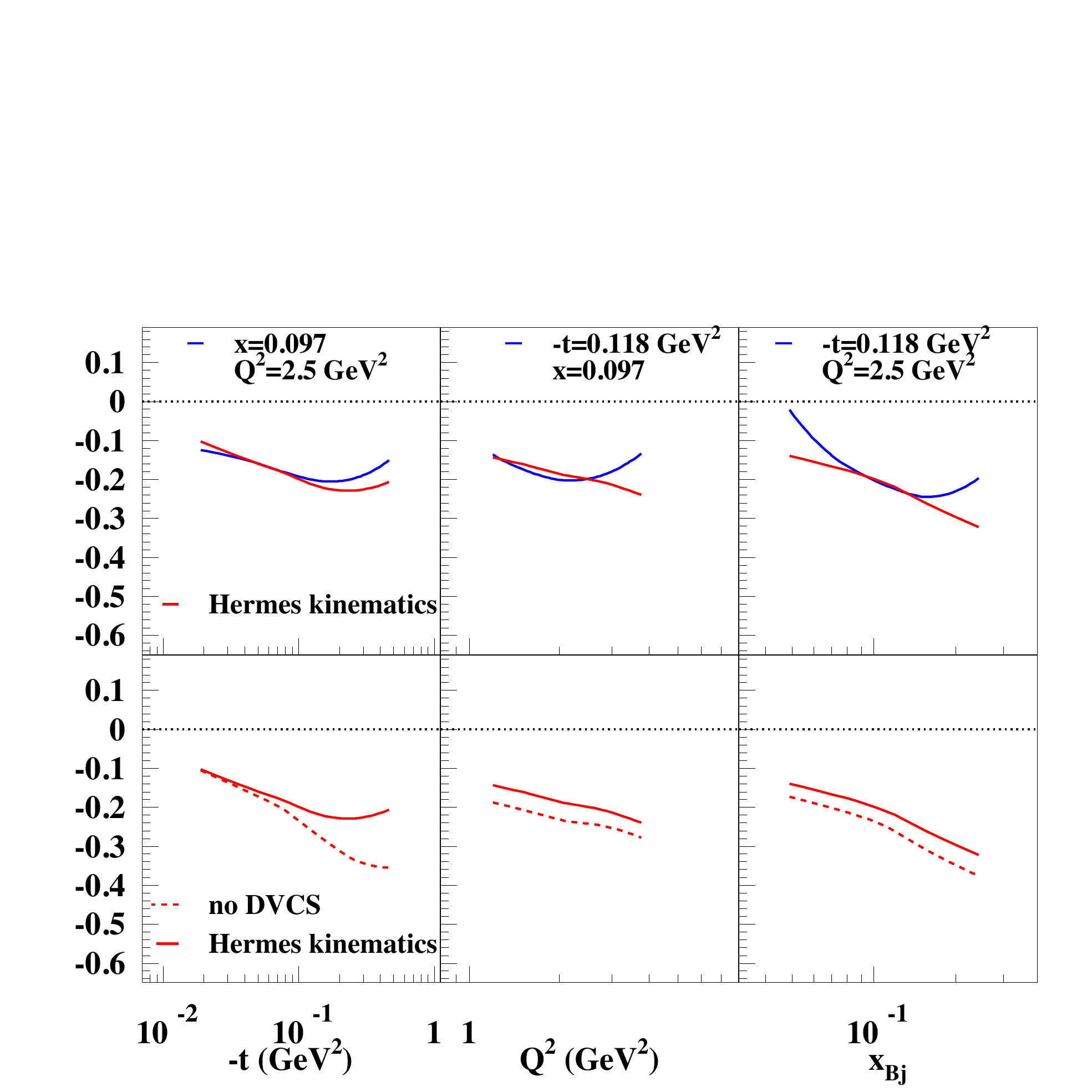}
\caption{(color online) Calculations at Hermes kinematics \cite{HERMES1,HERMES2,MM}. Shown is $A_{LU}(90^o)$ vs. $-t$, $Q^2$, and $x_{BJ}$, respectively, calculated at each kinematical bin provided by Hermes \cite{MM} (curve denoted as "Hermes kinematics"), and at 
the nominal average values presented in each panel. It is interesting to notice that due to the correlation between $x_{Bj}$ and $Q^2$ in the data, different features arise when using the average bin values. In the lower panels we also show the effect of disregarding the DVCS term in the denominator (dashed curves).}
\label{figH1}
\end{figure}
Hermes data are provided as "coefficients" of the azymuthal angles dependent terms. The dependence  of these
coefficients on the various kinematical variables is sensitive to the set of approximations that one uses in the extraction, thus 
affecting quantitative analyses. In order to facilitate the comparison, and to once more show some of the subtleties involved, in Fig.\ref{figH1} we show the results of our fit for  $A_LU$ vs. $-t$, $Q^2$, and $x_{Bj}$, respectively, calculated at each kinematical bin provided by Hermes \cite{MM} (curve denoted as "Hermes kinematics"), and at 
the nominal average values presented in each panel. It is interesting to notice that due to the correlation between $x_{Bj}$ and $Q^2$ in the data, different features arise when using the average bin values. In the figure (lower panels) we also show the effect of disregarding the DVCS term in the denominator. Similarly to the Jlab results, the GPD that the data are largely sensitive to is $H$, the role of the other GPDs being marginal. As a concluding remark on $A_{LU}$, we notice that Jefferson Lab Hall B data seem to suggest a decrease of $A_{LU}$ with $x_{Bj}$. The curves in Fig.\ref{figH1} used Jlab data in the fit, and therefore they show a definite slope in $x_{Bj}$. 

\begin{figure}
\includegraphics[width=9.5cm]{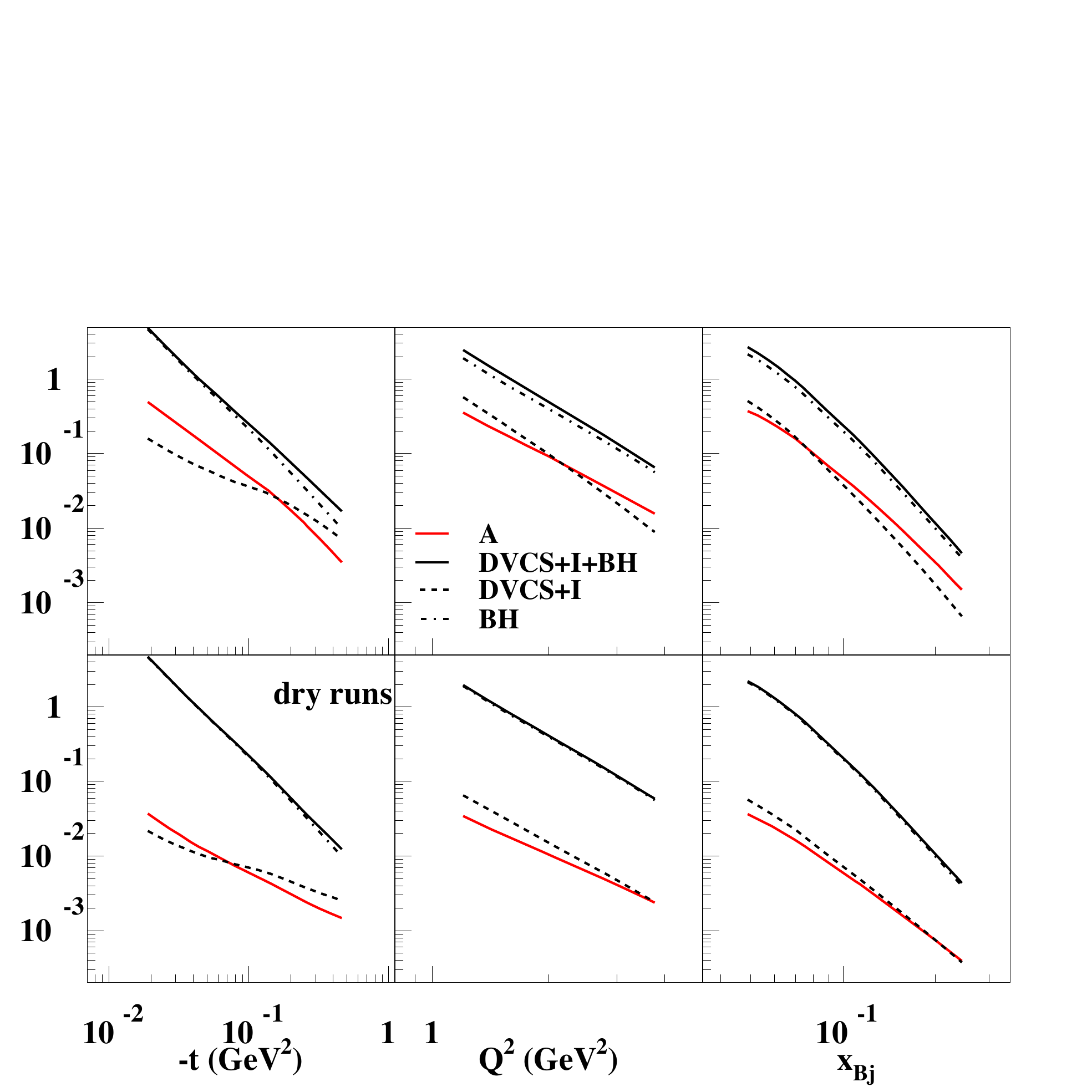}
\caption{(color online) Cross section components contributing to $A_{LU}(90^o)$: $A$, BH (dot-dashed lines), the sum of the $T_{DVCS}^2$ and 
the BH/DVCS interference terms (dashes), and $B$, Eqs.(\ref{ABC}).  The curves in the figure were calculated for the same kinematical bins as in Fig.\ref{figH1} \cite{MM}. 
In order to discern the role of the various components involving GPDs from the kinematics, we also show "dry runs" of our code in the lower panels obtained 
by setting all GPD factors equal to one.
}
\label{figH2}
\end{figure}
In Fig.\ref{figH2} we show the different cross section components contributing to $A_{LU}(90^o)$, given by $A$, and by the contributions from BH, $T_{DVCS}^2$, and 
the BH/DVCS interference terms in $B$, Eqs(\ref{ABC}).  The curves in the figure were calculated for the same kinematical bins as in Fig.\ref{figH1} \cite{MM}. 
In order to discern the role of the various components involving GPDs from the kinematics, we also show "dry runs" of our code in the lower panels obtained 
by setting all GPD factors equal to one. The various kinematical coefficients used in this analysis are written in Appendix \ref{appD}. 

\begin{figure}
\includegraphics[width=10.cm]{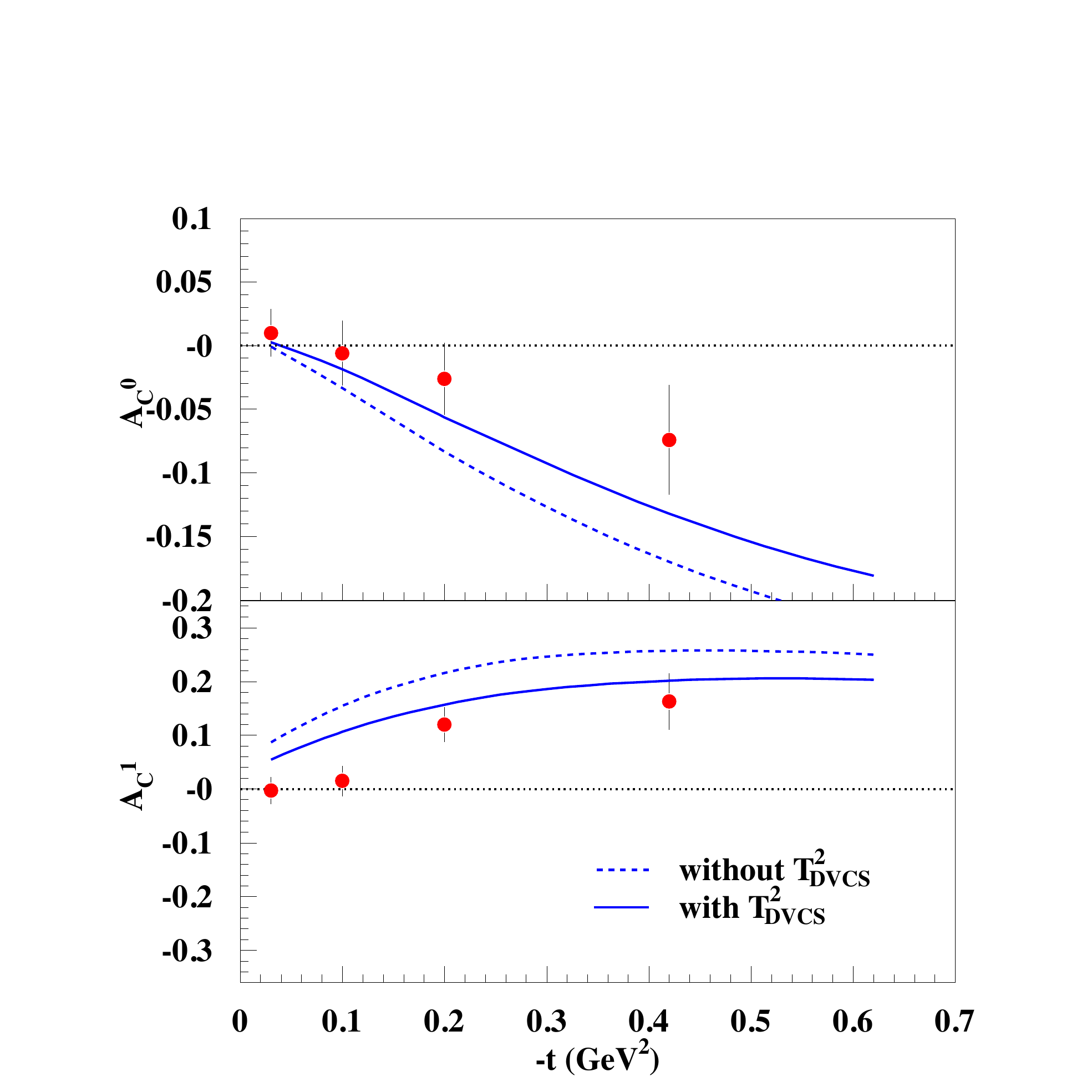}
\caption{Coefficients of the beam charge asymmetry, $A_C$, extracted from experiment \cite{HERMES1,HERMES2}. The lower panel is the coefficient for the 
$\cos \phi$ dependent term in Eq.(\ref{AC1}), while the upper panel is the $\cos \phi$ independent term. Notice the relevance of the pure DVCS contribution,
$\mid T_{DVCS} \mid^2$. }
\label{fig14}
\end{figure}
\begin{figure}
\includegraphics[width=10.cm]{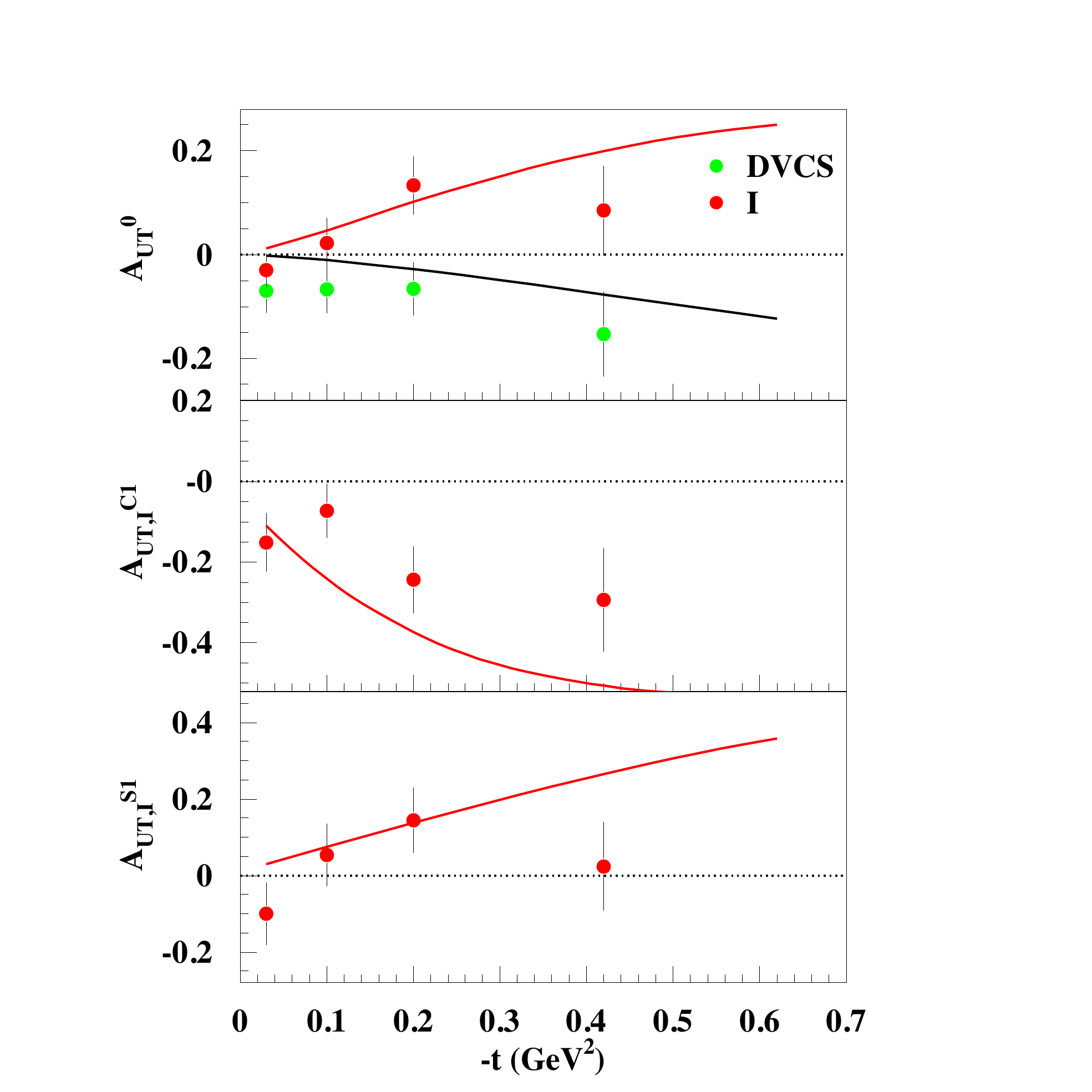}
\caption{Coefficients of the beam charge asymmetry, $A_{UT}$, extracted from experiment \cite{HERMES1,HERMES2}. The upper panel shows the terms
$E$ and $F$ from Eqs.(\ref{AUTD}) and (\ref{AUTI}), respectively; the middle panel shows $G$, and the lower panel $H$, both in Eq.(\ref{AUTI}). The curves are predictions obtained extending our quantitative
fit of Jefferson lab data to the Hermes set of observables.}
\label{fig15}
\end{figure}
Finally, in Fig.\ref{fig14} and Fig.\ref{fig15} our predictions for $A_C$ and $A_{UT}$ are shown vs. $-t$. The agreement with Hermes data is remarkably good within our theoretical error, despite we did not implement directly such data in the fit. 

\section{Conclusions and Outlook}
\label{sec5}
In this paper we have presented a parametrization of the chiral even GPDs that is inspired by a physically motivated picture of the nucleon as a quark-diquark system with Regge behavior. The spin structure of each of the four GPDs is determined via the covariant quark-nucleon scattering amplitude, with a diquark exchange. The masses, couplings and Regge power behavior that set the scale for the dependence on the kinematic variables, $X, \zeta, t, Q^2$, are determined via a recursive procedure. We fit the parton distribution functions $f_1$ and $g_1$ for the u and d quarks with $H(X,0,0)$ and $\widetilde{H}(X,0,0)$ at a low scale. The electromagnetic form factors, $F_1(t)$ and $F_2(t)$ constrain the first $X$ moments of $H(X,\zeta, t)$ and $E(X,\zeta,t)$. These first moments are constrained to satisfy polynomiality, thereby removing the $\zeta$ dependence and leaving only the $t$ dependence. This requires fixing the parameterization of the ERBL region, $X<\zeta$, for all $\zeta$ so as to satisfy a sum rule for the form factor. The same approach is used for the axial vector form factor and 
$\widetilde{H}$. Similarly the pseudoscalar form factor constrains 
$\widetilde{E}$, although we have not used that here (the contribution that is not dominated by the $\pi$ pole is not well known and is the subject of a forthcoming paper~\cite{GGL_Etilde}). 

In a previous paper Goldstein and Liuti~\cite{GL_partons} have shown that the simple parton interpretation of the ERBL region is dubious, so the parameterization used here for that region is chosen to have a polynomial form and to satisfy the proper crossing symmetry while maintaining polynomiality for the first moments. This constrains the $X-\zeta$ dependence through the sum rule Eq.\ref{FpERBL}. Having fixed the parameters of the Regge and diquark functions (Table I), the set of measured DVCS observables can be determined using evolution equations to match the $Q^2$ of different data sets. From these GPDs, the Compton Form Factors that enter the cross sections and asymmetries, can be computed. Beam asymmetry data indicate the need for a damping of the higher $\zeta$ behavior of the contributing CFFs. We incorporate this effect through a multiplicative function of $X, \zeta, t$ that lowers the high $\zeta$ value at higher $t$ (Eq.\ref{regge}). 

The final determination of the parameters provides an excellent fit to all of the available DVCS data. Newer DVCS cross section and asymmetry measurements at Jefferson Lab and COMPASS will provide a test of the flexibility of the model developed here. At this point we see that this physically motivated model provides a far reaching interpretation of the separate spin-dependent GPDs and thereby, a picture of the transverse structure of the nucleons will emerge. 

A number of questions remain that are being addressed in ongoing work. The connection between the Regge-like behavior of these GPDs and the more general form of variable mass diquark exchanges has opened up the possibility of having the Regge behavior emerge from diquark mass variation. Such a variation will better approximate the Fock space structure of the nucleon. A second, important concern is the inclusion of sea quarks, whose contribution will affect the low $x_{Bj}$ dependence, particularly the singlet, crossing even GPDs, whose Regge behavior is dominated by Pomeron exchange. Finally, the important extension of this parameterization scheme to the chiral odd GPDs is critical for the phenomenology of Deeply Virtual Meson Electroproduction, which was begun particularly for the $\pi^0$ in Ref.~\cite{AGL}. The connection of chiral odd GPDs to the transversity structure of the nucleon is of great interest as a signal of quark and gluon orbital angular momentum.

\acknowledgments
We thank H. Moutarde for useful exchanges, and for providing us with all available datasets. We are also indebted to Morgan Murray and Tanja Horn for 
comments and suggestions, and to Saeed Ahmad 
for participating in the initial stages of this work.
This work is supported by the U.S. Department
of Energy grants DE-FG02-01ER4120 (J.O.G.H., S.L.), and DE-FG02-92ER40702  (G.R.G.).

\vspace{0.3cm} 
\appendix 
\section{Useful Integration Formulae}
\label{appA}
A crucial point regarding the $\Delta_T$ dependence in Eqs.(\ref{GPDH},\ref{GPDE},\ref{GPDHTILDE},\ref{GPDETILDE}) is that the integral over ${\bf k}_\perp$ can be done explicitly over the azymuthal angle first, to yield the angular dependence. This is clear in noting that $k_1+i k_2 = |{\bf k}_\perp| e^{i\phi}$ and $\tilde{{\bf k}}_1+i \tilde{{\bf k}}_2=|{\bf k}_\perp| e^{i\phi}- (1-X)/(1-\zeta) {\bf \Delta_1}$, where  $\Delta_\perp$ can be chosen to be in the x-direction with no loss of generality. Also $\Delta_\perp \cdot {\bf k}_\perp = \Delta_1 |{\bf k}_\perp| cos\phi$. The $\phi$ dependence comes only from the ${\bf k}_\perp$ in the helicity flip numerators and the $k^{\prime 2}=k^2+\Delta^2-k^+\Delta^- + k^-\Delta^+ -2 \Delta_\perp \cdot {\bf k}_\perp$ in the denominators. 
When doing the integral over $\phi$ from $0$ to $2\pi$ the single factor $e^{i \phi}$gives 0, as does $cos\phi$ or $sin\phi$ alone. Hence single flip amplitudes will begin at $\Delta_\perp$ terms and double flip ones at $\Delta_\perp^2$.
The integrals used are of the form
\begin{subequations}
\begin{eqnarray}
\int\limits_0^{2 \pi} \frac{1}{(a - b \cos \phi)^2} \, d \phi = \frac{2 \pi a}{(a^2-b^2)^{3/2}}\\
\int\limits_0^{2 \pi} \frac{\cos \phi}{(a - b \cos \phi)^2} \, d \phi = \frac{2 \pi b}{(a^2-b^2)^{3/2}} 
\end{eqnarray}
\end{subequations}
where $b= 2 k_\perp \Delta_\perp$, and $a$ does not depend on $\Delta_\perp$.

\section{Initial scale generalized parton distributions}
\label{appB}
We present here a more practical version of the diquark contribution to the GPDs at the initial scale, $Q^2 = 0.0936$ GeV$^2$.
The complete parametrization needs to be multiplied by the Regge term provided in Eq.(\ref{regge}).
By defining 
\begin{subequations}
\begin{eqnarray}
L^2(X) & = & XM_{X}^{2} + (1-X)M_{\Lambda}^{2} -X(1-X)M^2
\label{def00}
\\
\mu  & = & m     +    XM
\label{def0}
\\
\mu' & = & m     +    X'M
\label{def1}
\\
X' & = & \frac{X-\zeta}{1-\zeta}
\label{def2}
\\
A & = & [L^2(X') + (1-X')^2 {\bf \Delta}_{\perp}^2]^2
\label{def3}
\\
B & = & 2[L^2(X') - (1-X')^2 {\bf \Delta}_{\perp}^2]
\label{def4}
\end{eqnarray}
\end{subequations}
all four GPDs can be written as follows,
\begin{align}
\label{GPD}
F(X,\zeta,t) = & \, \pi \, {\cal G}_1 \, (1-X)^{3/2}(1-X')^{3/2} \nonumber  \\ &\times \int\limits_{0}^{\infty}\frac{d\kappa}{[\kappa + L^2(X)]^2} \frac{g_0 +  g_1 \kappa +g_2 \kappa^2  }{[A + B\kappa + \kappa^2  ]^{3/2}} + {\cal G}_2,
\end{align}
where for $E$
\begin{subequations}
\begin{eqnarray}
{\cal G}_1 & = & 2(1-X')\sqrt{1-\zeta} R(X,\zeta,t)
\label{def5}
\\
{\cal G}_2 & = & 0
\label{def6}
\\
g_0 & = & \mu M [L^2(X')+(1-X')^2{\bf \Delta}_{\perp}^2]
\label{def7}
\\
g_1 & = & -M(\mu-2\mu')
\label{def8}
\\
g_2 & = & 0
\label{def9}
\end{eqnarray}
\end{subequations}
For $\widetilde{E}$,
\begin{subequations}
\begin{eqnarray}
{\cal G}_1 & = & 2(1-X')\sqrt{1-\zeta} \left( 2 \, \frac{1-\frac{\zeta}{2}}{\zeta} \right)
\label{def10}
\\
{\cal G}_2 & = & 0
\label{def11}
\\
g_0 & = & \mu M[L^2(X')+(1-X')^2{\bf \Delta}_{\perp}^2]
\label{def12}
\\
g_1 & = & -M(\mu+2\mu')
\label{def13}
\\
g_2 & = & 0
\label{def14}
\end{eqnarray}
\end{subequations}
For $H$,
\begin{subequations}
\begin{eqnarray}
{\cal G}_1 & = & \frac{1-\frac{\zeta}{2}}{\sqrt{1-\zeta}}
\label{def15}
\\
{\cal G}_2 & = & \frac{\zeta^2}{4(1-\zeta)} E
\label{def16}
\\
g_0 & = & \mu \mu' [L^2(X')+(1-X')^2{\bf \Delta}_{\perp}^2]
\label{def17}
\\
g_1 & = & \mu \mu' + [L^2(X')-(1-X')^2{\bf \Delta}_{\perp}^2]
\label{def18}
\\
g_2 & = & 1
\label{def19}
\end{eqnarray}
\end{subequations}
For $\widetilde{H}$,
\begin{subequations}
\begin{eqnarray}
{\cal G}_1 & = & \frac{1-\frac{\zeta}{2}}{\sqrt{1-\zeta}}
\label{def20}
\\
{\cal G}_2 & = & \frac{\zeta^2}{4(1-\zeta)}\widetilde{E}
\label{def21}
\\
g_0 & = & \mu \mu' [L^2(X')+(1-X')^2{\bf \Delta}_{\perp}^2]
\label{def22}
\\
g_1 & = & \mu \mu' - [L^2(X')-(1-X')^2{\bf \Delta}_{\perp}^2]
\label{def23}
\\
g_2 & = & -1
\label{def24}
\\  \nonumber
\end{eqnarray}
\end{subequations}
In the forward limit the GPD's reduce to the form
\begin{align}
\label{GPD_forward}
G = \, \pi \, {\cal G}_1 (1-X)^3  \, \int\limits_{0}^{\infty}d\kappa \frac{g_0 +  g_1 \kappa +g_2 \kappa^2  }{[\kappa + L^2(X)]^5}
\end{align}
where the functions $g_0, g_1, g_2$ are evaluated at ${\bf \Delta}_{\perp}^2=0$ and $ \zeta=0$. Solving the integral for each case we get

\begin{eqnarray}
\label{H_forward}
H(X,0,0) & =  & (1-X)^3 \, \frac{ 2\mu^2 + L^2(X) }{6L^6} X^{-\alpha}
\\
\label{Htilde_forward}
\widetilde{H}(XA,0,0) & = &  \pi \, (1-X)^3 \frac{ 2\mu^2 - L^2(X) }{6L^6}  X^{-\alpha}
\\
\label{E_forward}
E(X,0,0) & = & 2  (1-X)^4  \frac{\mu^2 M}{3L^6} X^{-\alpha} 
\\
\label{Etilde_forward}
\widetilde{E}(X,0,0) & =  & \pi M (1-X)^6 X^{-\alpha} \left[\frac{M}{3L^6} \right.
\\
 & - & \left.\frac{4  \mu\left[(1-2X)M^2-M_X^2+M_\Lambda^2\right]}{5L^8}\right] 
\end{eqnarray}

\section{Principal Value Integration}
The PV integrations entering the CFFs defined in Eq.(\ref{Re_H}) were carried out with the modified Gaussian method, yielding 
\begin{eqnarray}
& & \Re e {\cal H}  =  PV \int\limits_{\zeta/2}^1 dX \frac{H^+(X,\zeta,t)}{X-\zeta} + \int\limits_{\zeta/2}^1 dX \frac{H^+(X,\zeta,t)}{X} \nonumber \\
& = & H(\zeta,\zeta,t) \ln \frac{ 1 - \zeta}{ \zeta/2} + \int\limits_{\zeta/2}^{1} dX \frac{H^+(X,\zeta,t) - H(\zeta,\zeta,t)}{X-\zeta}  \nonumber
 \\
& + &  \int\limits_{\zeta/2}^1 dX \frac{H^+(X,\zeta,t)}{X} 
\end{eqnarray}

\section{Kinematical coefficients in asymmetries and cross sections}
\label{appD}
The kinematical coefficients in Eqs.(\ref{ABC}) are
\begin{eqnarray}
K_I(\phi) & = & \frac{-8 K (2-y)}{x_{Bj} y^2 P_1(\phi) P_2(\phi) t}  \\
K_{DVCS}^0 & = &  \frac{1}{y^2 Q^2} \frac{2 (2-2y+y^2)}{(2-x_{Bj})^2} 4 (1-x_{Bj}) \\
K_I^0 & =  &  \frac{1}{x_{Bj} y^3 P_1(\phi) P_2(\phi) t}  \frac{8(2-y)^3}{1-y} K^2
\end{eqnarray}
with \cite{BKM}
\begin{eqnarray}
P_1(\phi) & = &  - \frac{1}{y \sqrt{1+ \epsilon^2}} ( H +  2 K \cos\phi )  \\
P_2(\phi) & = & \left(1+\frac{t}{Q^2}\right) - P_1(\phi),
\end{eqnarray}
where
\[ H=   \left(1- y - \frac{1}{2}y \epsilon^2\right) \left(1+\frac{t}{Q^2} \right) - (1-x_{Bj})(2-y)\frac{t}{Q^2}, \] 
and $K^2 \propto -t/Q^2$.  


\end{document}